\documentclass[twocolumn]{emulateapj}
\begin{document}
\newcommand{\msun}{\mbox{M$_{\odot}$}}
\newcommand{\rsun}{\mbox{R$_{\odot}$}}
\newcommand{\zsun}{\mbox{Z$_{\odot}$}}
\newcommand{\lsun}{\mbox{L$_{\odot}$}} 

\title{{\it Spitzer} SAGE-SMC Infrared Photometry of Massive Stars \\in
  the Small Magellanic Cloud}

\author{A.\,Z. Bonanos\altaffilmark{1}, D.\,J. Lennon\altaffilmark{2,3},
  F. K\"ohlinger\altaffilmark{4}, J.\,Th.\ van Loon\altaffilmark{5},
  D.\,L. Massa\altaffilmark{2}, M. Sewilo\altaffilmark{2},
  C.\,J. Evans\altaffilmark{6}, N. Panagia\altaffilmark{2,7,8},
  B.\,L. Babler\altaffilmark{9}, M. Block\altaffilmark{10},
  S. Bracker\altaffilmark{9}, C.\,W. Engelbracht\altaffilmark{9},
  K.\,D. Gordon\altaffilmark{2}, J.\,L. Hora\altaffilmark{11},
  R. Indebetouw\altaffilmark{12}, M.\,R. Meade\altaffilmark{9},
  M. Meixner\altaffilmark{2}, K.\,A. Misselt\altaffilmark{10},
  T.\,P. Robitaille\altaffilmark{11}, B. Shiao\altaffilmark{2},
  B.\,A. Whitney\altaffilmark{13}}

\altaffiltext{1}{Institute of Astronomy \& Astrophysics, National
  Observatory of Athens, I. Metaxa \& Vas. Pavlou St., P. Penteli, 15236
  Athens, Greece; bonanos@astro.noa.gr} \altaffiltext{2}{Space Telescope
  Science Institute, 3700 San Martin Drive, Baltimore, MD, 21218, USA;
  lennon, massa, sewilo, panagia, kgordon, meixner, shiao@stsci.edu}
\altaffiltext{3}{European Space Agency, Research and Scientific Support
  Department, Baltimore, MD 21218, USA} \altaffiltext{4}{Department of
  Physics and Astronomy, Heidelberg University, Albert-Ueberle-Str. 3-5,
  D-69120 Heidelberg, Germany; koehlinger@stud.uni-heidelberg.de}
\altaffiltext{5}{Astrophysics Group, Lennard-Jones Laboratories, Keele
  University, Staffordshire ST5 5BG, UK; jacco@astro.keele.ac.uk}
\altaffiltext{6}{UK Astronomy Technology Centre, Royal Observatory
  Edinburgh, Blackford Hill, Edinburgh, EH9 3HJ, UK;
  chris.evans@stfc.ac.uk} \altaffiltext{7}{INAF/Osservatorio Astrofisico
  di Catania, Via Santa Sofia 78, I-95123 Catania, Italy}
\altaffiltext{8}{Supernova Ltd., OYV \#131, Northsound Road, Virgin
  Gorda, British Virgin Islands} \altaffiltext{9}{Department of
  Astronomy, 475 North Charter Street, University of Wisconsin, Madison,
  WI 53706, USA; brian@sal.wisc.edu, s\_bracker@hotmail.com,
  meade@astro.wisc.edu} \altaffiltext{10}{Steward Observatory,
  University of Arizona, 933 North Cherry Avenue, Tucson, AZ 85721, USA;
  mblock, cengelbracht, kmisselt@as.arizona.edu}
\altaffiltext{11}{Harvard-Smithsonian Center for Astrophysics, 60 Garden
  Street, MS 67, Cambridge, MA 02138, USA; jhora,
  trobitaille@cfa.harvard.edu} \altaffiltext{12}{Department of
  Astronomy, University of Virginia, PO Box 3818, Charlottesville, VA
  22903, USA; remy@virginia.edu} \altaffiltext{13}{Space Science
  Institute, 4750 Walnut Street, Suite 205, Boulder, CO 80301, USA;
  bwhitney@spacescience.org}

\begin{abstract}

We present a catalog of 5324 massive stars in the Small Magellanic Cloud
(SMC), with accurate spectral types compiled from the literature, and a
photometric catalog for a subset of 3654 of these stars, with the goal
of exploring their infrared properties. The photometric catalog consists
of stars with infrared counterparts in the {\it Spitzer}\, SAGE-SMC
survey database, for which we present uniform photometry from $0.3-24$
$\mu$m in the $UBVIJHK_{s}$+IRAC+MIPS24 bands. We compare the color
magnitude diagrams and color-color diagrams to those of stars in the
Large Magellanic Cloud (LMC), finding that the brightest infrared
sources in the SMC are also the red supergiants, supergiant B[e]
(sgB[e]) stars, luminous blue variables, and Wolf-Rayet stars, with the
latter exhibiting less infrared excess, the red supergiants being less
dusty and the sgB[e] stars being on average less luminous. Among the
objects detected at 24~$\mu$m in the SMC are a few very luminous
hypergiants, 4 B-type stars with peculiar, flat spectral energy
distributions, and all 3 known luminous blue variables. We detect a
distinct Be star sequence, displaced to the red, and suggest a novel
method of confirming Be star candidates photometrically. We find a
higher fraction of Oe and Be stars among O and early-B stars in our SMC
catalog, respectively, when compared to the LMC catalog, and that the
SMC Be stars occur at higher luminosities. We estimate mass-loss rates
for the red supergiants, confirming the correlation with luminosity even
at the metallicity of the SMC.  Finally, we confirm the new class of
stars displaying composite A \& F type spectra, the sgB[e] nature of
2dFS1804 and find the F0 supergiant 2dFS3528 to be a candidate luminous
blue variable with cold dust.

\end{abstract}


\section{Introduction}
\label{section:intro}

The {\em Spitzer Space Telescope} Legacy Surveys SAGE \citep[``Surveying
  the Agents of a Galaxy's Evolution'',][]{Meixner06} and SAGE-SMC
\citep{Gordon10} have for the first time made possible a comparative
study of the infrared properties of massive stars at a range of
metallicities, by imaging both the Large and Small Magellanic Clouds
(LMC and SMC). In \citet[][hereafter Paper I]{Bonanos09a}, we presented
infrared properties of massive stars in the LMC (at 0.5\,\zsun). The
motivation for that study was twofold: (a) to use the infrared excesses
of massive stars to probe their winds, circumstellar gas and dust, and
(b) to provide a template for studies of other, more distant,
galaxies. Paper I was the first major compilation of accurate spectral
types and multi-band photometry from 0.3$-$24 $\mu$m for massive stars
in any galaxy, increasing by an order of magnitude the number of massive
stars for which mid-infrared photometry was available. The recently
completed SAGE-SMC survey offers the opportunity to extend the study of
infrared properties of massive stars to a metallicity of approximately
0.2\,\zsun~\citep[see e.g.][]{Hunter07}, and enables the investigation
of their dependence on metallicity, at least over the range
0.2--0.5\,\zsun.

Infrared excess in hot massive stars is primarily due to free free
emission from their ionized, line-driven, stellar
winds. \citet{Panagia75} and \citet{Wright75} first computed the
free-free emission from ionized envelopes of hot massive stars, as a
function of the mass-loss rate ($\dot{M}$) and the terminal velocity of
the wind ($v_{\infty}$). The properties of massive stars, and in
particular their stellar winds (which affect their evolution) are
expected to depend on metallicity ($Z$). For example, \citet{Mokiem07a}
found empirically that mass-loss rates scale as $\dot{M} \sim Z^{0.83\pm
  0.16}$, in good agreement with theoretical predictions
\citep{Vink01}. The expectation, therefore, is that the infrared
excesses of OB stars in the SMC should be lower than in the LMC, given
that $\dot{M}$ is lower in the SMC\footnote{The dependence of
  $v_{\infty}$ on $Z$ is negligible for this range of $Z$
  \citep[see][]{Leitherer92, Evans04b}.}. Furthermore, there is strong
evidence that the fraction of classical Be stars among B-type stars is
higher at lower metallicity \citep[possibly due to faster rotation, as
  measured by][]{Martayan07b}.  \citet{Grebel92} were the first to find
evidence for this, by showing that the cluster NGC\,330 in the SMC has
the largest fraction of Be stars of any known cluster in the Galaxy, LMC
or SMC. More recent spectroscopic surveys \citep{Martayan10} have
reinforced this result. In Paper I, we showed that the Be stars in the
LMC are easily discriminated by their mid-infrared colors; we are
therefore interested in a comparison with the SMC to quantify the global
dependence of the Be star fraction on metallicity.  The incidence of
Be/X-ray binaries is also much higher in the SMC than in the LMC
\citep{Liu05}, while the incidence of Wolf-Rayet (WR) stars is much
lower; therefore, a comparison of infrared excesses for these objects is
also of interest. Finally, there is interest in the metallicity
dependence of the mid-infrared colors of red supergiants, which probe
circumstellar dust.

Following the same strategy as in Paper I, we compiled a catalog of
massive stars, which we cross-matched in the SAGE-SMC database to study
their infrared properties. This paper is organized as follows:
Section~\ref{sec:catalog} describes our spectroscopic and photometric
catalogs of massive stars in the SMC, Section~\ref{sec:cmd} presents the
resulting color--magnitude and two--color diagrams, and Section
\ref{sec:hotstars} presents the infrared excesses detected in various
types of massive stars. Section~\ref{sec:rsg_mdot} investigates the
mass-loss rates in red supergiants, and Section \ref{sec:summary}
summarizes our results.

\section{Catalog of Massive Stars in the SMC}
\label{sec:catalog}

As in Paper I, we have compiled a catalog of massive stars with known
spectral types in the SMC from the literature. We then cross-matched the
stars in the SAGE-SMC database, after incorporating optical and
near-infrared photometry from recent surveys of the SMC. The resulting
photometric catalog was used to study the infrared properties of the
stars. In Section \ref{sec:sptypecatalog} we describe the spectral-type
catalog compiled from the literature, in Section \ref{sec:oirsurveys}
the existing optical, near-infrared and mid-infrared surveys of the SMC,
which were included in the SAGE-SMC database, and in Section
\ref{sec:photcatalog} the cross-matching procedure and the resulting
photometric catalog.

\subsection{Spectral-type Catalog}
\label{sec:sptypecatalog}

The criteria for inclusion in the spectral type catalog of massive SMC
stars were the availability of accurate coordinates and accurate (mainly
optical) spectral classifications that correspond to stars with masses
$\gtrsim8\,\msun$. The 2dF Survey \citet[2dFS;][4161 objects]{Evans04}
comprises the largest catalog of SMC stars with spectroscopic
classifications and therefore makes up the backbone of our spectral type
catalog\footnote{By retaining all SMC stars from 2dFS, we are possibly
  including some stars with masses $<8\,\msun$.}. Previous compilations
\citep[e.g.][]{Azzopardi75, Azzopardi79, Azzopardi82} were included with
updated coordinates from Brian Skiff's
lists\footnote{ftp://ftp.lowell.edu/pub/bas/starcats/} and spectral
types from the literature, when available. The literature search
resulted in 5324 entries. We do not claim completeness; however, not
only have we targeted studies of hot massive stars (defined by their
spectral types) in individual clusters and OB associations, but also
studies of particular types of massive stars. The largest of these
include: the survey of B and Be stars in NGC\,330 \citep[][314
  stars]{Martayan07a, Martayan07b}, the VLT-FLAMES Survey of Massive
Stars in NGC\,330 and NGC\,346 \citep[][232 stars]{Evans06}, and
unpublished additional FLAMES observations in NGC\,346 (PI: Evans;
I. Hunter et al. 2010, in preparation, 244 OB \& A-type stars). We have
also added the 3 known luminous blue variables (LBVs), 5 supergiant B[e]
stars \citep{Zickgraf06}, 69 Be/X-ray binaries \citep[with spectral
  types from the updated online
  catalog\footnote{http://xray.sai.msu.ru/\textasciitilde
    raguzova/BeXcat/} of][]{Raguzova05}, early-type eclipsing and
spectroscopic binaries, and red supergiants \citep[][107
  stars]{Massey03b, Levesque06, Levesque07}. The completeness of the
catalog depends on the spectral type, e.g., it is $\sim4 \%$ for the
unevolved O stars in our catalog \citep[out of an estimated total of
  2800 unevolved stars with masses $>20\,\msun$ in the
  SMC;][]{Massey09}, while the WR catalog \citep{Massey03c} is thought
to be close to complete.

Table~\ref{tab:catalog} presents our catalog of 5324 massive stars,
sorted by R.A., listing: the star name(s), coordinates in degrees
(J2000.0), the reference and corresponding spectral classification. The
names of the stars are taken from the corresponding reference. The
spectral classifications in the catalog are typically accurate to one
spectral type and one luminosity class. We retained about a dozen stars
from the above studies with only approximate spectral types (e.g.,
``early B''). For double entries, we selected the most precise spectral
classification available, usually corresponding to the most recent
reference.

\subsection{Optical and Infrared Surveys of the SMC}
\label{sec:oirsurveys}

Several large optical and infrared photometric catalogs of the SMC have
recently become available, enabling us to obtain accurate photometry for
its massive star population in the wavelength range $0.3-8$~$\mu$m and
in some cases up to $24$~$\mu$m. The optical surveys are as follows: the
$UBVR$ catalog of \citet{Massey02} with 85,000 bright stars in an area
covering 7.2 square degrees, the $UBVI$ Magellanic Clouds Photometric
Survey \citep[MCPS;][]{Zaritsky02} including 5.1 million stars in the
central 18 square degrees of the SMC, and the OGLE III catalog
containing $VI$ photometry of 6.2 million stars covering 14 square
degrees \citep{Udalski08b}. The angular resolution is $\sim2\farcs6$ for
the catalog of \citet{Massey02}, $\sim1\farcs5$ for MCPS, and
$\sim1\farcs2$ for OGLE~III.

The existing near-infrared photometric catalogs include: the Two Micron
All Sky Survey \citep[2MASS;][extended by 6X2MASS]{Skrutskie06} and the
targeted IRSF survey \citep{Kato07}, which contains 2.8 million sources
in the central 11 square degrees of the SMC. 2MASS has a pixel scale of
$2\farcs0$~pixel$^{-1}$, an average seeing of $2\farcs5$ and limiting
magnitudes of $J=15.8$, $H=15.1$ and $K_s=14.3$. IRSF has a pixel scale
of $0\farcs45$~pixel$^{-1}$, average seeing of $1\farcs3$, $1\farcs2$,
$1\farcs1$ in the $JHK_s$ bands, respectively, and limiting magnitudes
of $J=18.8$, $H=17.8$ and $K_s=16.6$. In the mid-infrared, the {\it
  Spitzer} SAGE-SMC survey uniformly imaged the whole SMC ($\sim$30
square degrees) in the IRAC and MIPS bands on two epochs in 2007-2008
\citep{Gordon10}. The survey has recently produced a combined mosaic
catalog of 1.2 million sources. IRAC, with a pixel scale of
$1\farcs2$~pixel$^{-1}$, yields an angular resolution of
$1\farcs7-2\farcs0$ and MIPS at 24~$\mu$m has a resolution of
$6\arcsec$.

Given the variation in the depth, resolution and spatial coverage of
these surveys, we included the available photometry for the massive
stars in our catalog from all the MCPS, OGLE~III, 2MASS, IRSF and
SAGE-SMC catalogs. MCPS has incorporated the catalog of \citet{Massey02}
for bright stars common to both catalogs. Photometry of higher accuracy,
particularly in the optical, exists in the literature for many of the
stars in our catalog; however, it was not included in favor of
uniformity.

\subsection{Photometric Catalog}
\label{sec:photcatalog}

\subsubsection{Matching Procedure}

We used the following SAGE-SMC data products to search for the
mid-infrared counterparts to the 5324 massive stars listed in
Table~\ref{tab:catalog}: the v1.0 IRAC Catalog containing IRAC Epoch 1
data bandmerged with All-Sky 2MASS and 6X2MASS catalogs, and the MIPS 24
$\mu$m Epoch 1 Catalog. The details about the SAGE-SMC data processing
and Legacy data products can be found in \citet{Gordon10} and in ``The
SAGE-SMC Data Description: Delivery 1'' document available at the
NASA/IPAC Infrared Science Archive Web
site\footnote{http://irsa.ipac.caltech.edu/data/SPITZER/SAGE/}.  The
IRAC Catalog and MIPS 24 $\mu$m Catalog contain high-reliability sources
and are subsets of more complete, but less reliable source lists: the
IRAC Archive and MIPS 24 $\mu$m full list (see the data delivery
document for a comparison between the catalogs and archives/full lists).

We started constructing the photometric catalog for the massive stars in
our list by selecting their SAGE-SMC IRAC Epoch 1 Catalog
counterparts. We performed a conservative neighbor search with a 1$''$
search radius, and selected the closest match for each source. We found
mid-infrared counterparts for 3654 out of 5324 sources. The IRAC Epoch 1
catalog (IRACC), MIPS 24 $\mu$m catalog, IRSF, MCPS, and OGLE III
catalogs were crossed-matched in the SAGE-SMC
database\footnote{http://mastweb.stsci.edu/scasjobs/} to provide
photometry for sources over a wavelength range from 0.3-24 $\mu$m. We
used this ``universal catalog'' to extract multi-wavelength photometry
for IRAC sources matched to the massive stars.  Specifically, for IRAC
sources with one or more matches in other catalogs (all but 5), we only
considered the closest matches between sources from any two available
catalogs (IRAC-MIPS24, IRAC-MCPS, MIPS24-MCPS, IRAC-IRSF, MIPS24-IRSF,
IRAC-OGLE III, MIPS24-OGLE III), with distances between the matched
sources of $\leq$1$''$. For example, for a match between the IRAC, MIPS
24 $\mu$m and MCPS catalogs (IRAC-MIPS24-MCPS), we applied these
constraints on the IRAC-MIPS24, MIPS24-MCPS, and IRAC-MCPS matches. If
the match was not the closest one or the distance was $>1''$, we dropped
the pair. These stringent criteria were used to ensure that sources from
individual catalogs for each multi-catalog match refer to the same
star. Table~\ref{tab:matchtype} shows the breakdown of the matched stars
to the catalogs, such that 40 stars were only matched to the IRACC, 214
only to the IRACC+IRSF catalogs etc.
 
\subsubsection{Catalog Description}

Table~\ref{tab:phot} presents our final matched catalog of 3654 stars,
with the star name, IRAC designation, $UBVIJHK_{s}$+IRAC+MIPS24
photometry and errors, reference paper, corresponding spectral
classification and comments, sorted by increasing R.A.. Overall, the
photometry is presented in order of shortest to longest wavelength. The
17 columns of photometry are presented in the following order: $UBVI$
from MCPS, $VI$ from OGLE~III, $JHK_s$ from 2MASS, $JHK_s$ from IRSF,
IRAC 3.6, 4.5, 5.8, 8.0 $\mu$m (or [3.6], [4.5], [5.8], [8.0]) and MIPS
24 $\mu$m (or [24]). A column with the associated error follows each
measurement, except for the $VI$ photometry from OGLE~III. Henceforth,
$JHK_s$ magnitudes refer to 2MASS photometry, whereas IRSF photometry is
denoted by a subscript, e.g.\ $J_{IRSF}$. All magnitudes are calibrated
relative to Vega \citep[e.g.\ see][for IRAC]{Reach05}. In
Table~\ref{tab:filters}, we summarize the characteristics of each
filter: the effective wavelength $\lambda_{\rm eff}$, zero magnitude
flux (in Jy), angular resolution, and the number of detected stars in
each filter.

\begin{figure*}[ht]  
\includegraphics[angle=270,width=7.in]{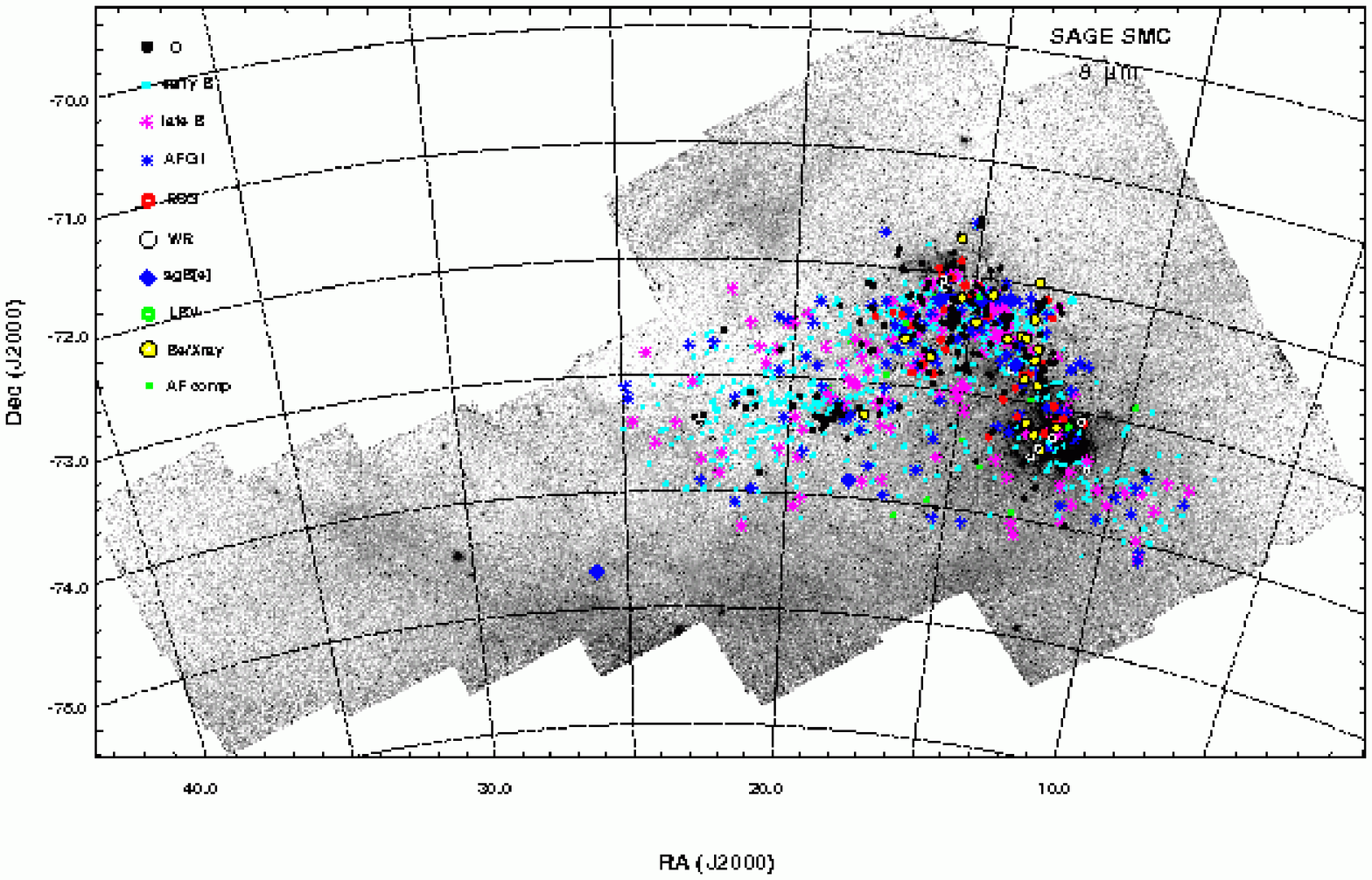}
\caption{Spatial distribution of massive stars with IRAC counterparts,
  overplotted onto the 8~$\mu$m SAGE image (Right Ascension and
  Declination are given in degrees). Different symbols denote different
  spectral types. For clarity, only a third of the early and late B
  stars are shown; these are equally distributed along the SMC bar and
  wing. Other types are concentrated along the bar, due to the fact that
  most spectroscopic surveys have targeted the bar.}
\label{spatial}
\end{figure*}

The spatial distribution of our 3654 matched sources is shown in
Figure~\ref{spatial}, overlaid onto the 8~$\mu$m image of the SMC. For
clarity, only a third of the early and late-B stars are shown; these are
equally distributed along the SMC bar and wing. Other types are
concentrated along the bar, partly due to the fact that most
spectroscopic surveys have targeted the bar. The 8~$\mu$m emission,
which maps the surface density of the interstellar medium, is stronger
in the SMC bar than the wing, since the former contains more
star-forming regions.

\section{Infrared Color--Magnitude and Two--Color Diagrams}
\label{sec:cmd}

We divide the matched stars into 10 categories according to their
spectral types: O stars, early (B0$-$B2.5) and late (B3$-$B9) B stars
(the majority of these have supergiant or giant luminosity
classifications), spectral type A, F and G type (AFG) supergiants, K and
M red supergiants (RSGs), WR stars, supergiant B[e] (sgB[e]) stars,
confirmed LBVs, Be/X-ray binaries and stars with composite features of
both A and F stars \citep[AFcomp;][see also
  Section~\ref{sub:afcomp}]{Evans04}. Note, that 813 stars not included
in any of the above categories are classified as ``other'' (and only
shown in certain plots)\footnote{Note that 5 Bextr, 5 Be, 1 Be(FeII), 1
  Bpec, and 1 A7-F5e star are found among these.}. Most are A stars with
luminosity class II or F/G stars without a luminosity classification
from \citet{Evans04}, although with $v_{radial}>100$~km~s$^{-1}$. To
allow for a direct comparison with the LMC results of Paper I, we
present the same color--magnitude and two--color diagrams (CMDs and
TCDs), identifying stars in the 10 categories described
above.\footnote{The conversion to absolute magnitudes in all CMDs is
  based on a true SMC distance modulus of 18.91 mag \citep{Hilditch05}.}

\begin{figure}[ht]  
\includegraphics[width=3.5in]{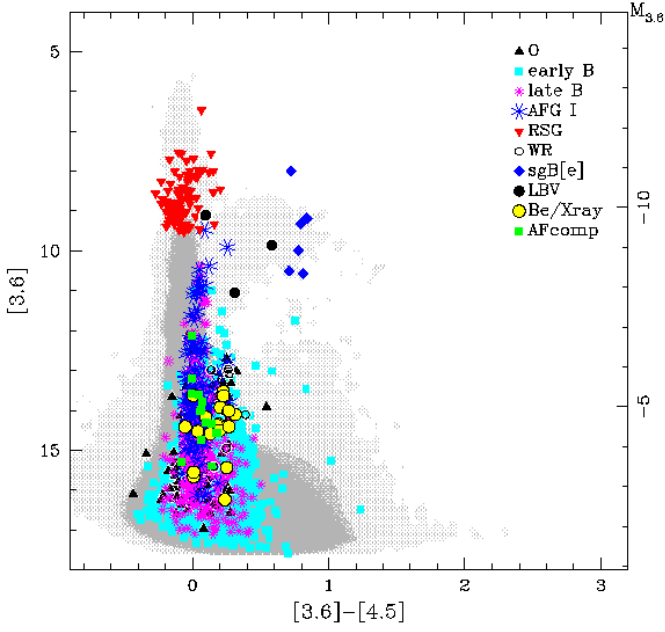}
\caption{[3.6] versus $[3.6]-[4.5]$ color magnitude diagram for massive
  stars with IRAC counterparts in the SAGE database. The conversion to
  absolute magnitudes is based on a true SMC distance modulus of 18.91
  mag \citep{Hilditch05}. Different symbols denote different spectral
  types. The locations of all the SAGE detections are shown in gray as a
  Hess diagram. The RSGs, sgB[e] and LBVs are among the most luminous
  stars at 3.6~$\mu$m.}
\label{cmd36}
\end{figure}

\begin{figure}[ht]  
\includegraphics[width=3.5in]{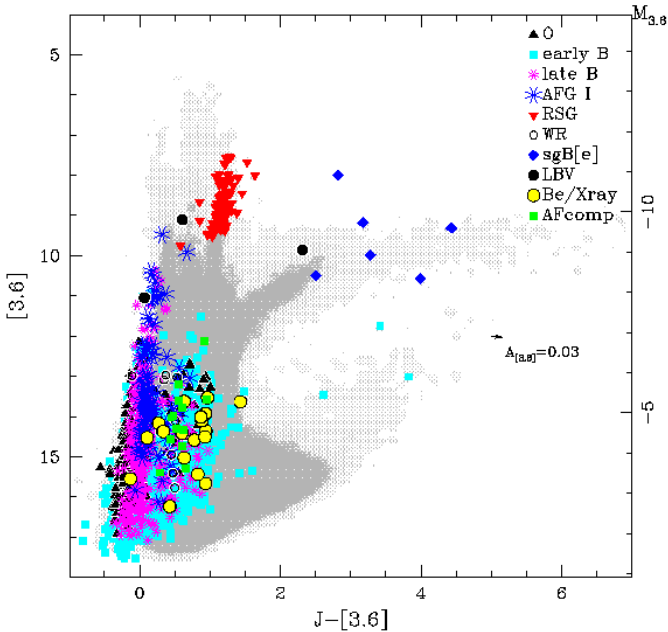}
\caption{Same as Figure~\ref{cmd36}, but for the [3.6] versus $J-[3.6]$
  color magnitude diagram. The reddening vector for $E(B-V)=0.2$ mag is
  shown. The longer baseline separates the populations more clearly.}
\label{cmd36j36}
\end{figure}

Infrared $[3.6]$ versus $[3.6]-[4.5]$ and $J-[3.6]$ CMDs for all the
stars in the catalog are shown in Figures~\ref{cmd36} and
\ref{cmd36j36}. The locations of all the SAGE-SMC catalog detections are
represented by the gray two dimensional histogram (Hess diagram). The
red giants form the clump at $[3.6]>15$ mag, while the vertical blue
extension contains late-type SMC and foreground stars (free--free
emission causes the OB stars to have redder colors). The asymptotic
giant branch stars are located at $[3.6]\sim11$~mag, and
$J-[3.6]\sim2$~mag. The RSGs, sgB[e], and LBVs are among the brightest
stars at 3.6 $\mu$m and occupy distinct regions in the diagrams (as in
the LMC). Most of the O and B stars are located along a vertical line at
$[3.6]-[4.5]\sim0$, as expected. The RSGs have ``blue'' colors because
of the depression at [4.5] due to the CO band \citep[see
  e.g.][]{Verhoelst09}. The sgB[e] stars are 0.6-0.8 mag redder and can
have similar brightnesses to the RSGs. The late-B stars are brighter
than the early-B stars because most of the former are luminous
supergiants. A reddening vector for $E(B-V)=0.2$ mag
\citep[corresponding to some of the largest values for color excess
  found by][]{Massey95, Larsen00} is shown in Figure~\ref{cmd36j36} to
illustrate the small effect of reddening, which decreases at longer
wavelengths. We note that the reddest of the bright AFG supergiants is
the F0I eclipsing binary R47 \citep{Prieto08c}. In
Figure~\ref{cmd36j36}, we find 3 early B-type stars with colors similar
to the sgB[e] stars, although they are fainter at [3.6]. In order of
increasing $J-[3.6]$ color, these stars are: 2dFS0402, 2dFS2411 and
2dFS2673 (all classified as B0-5V). Their evolutionary status is not
clear, though they are worthy of future study (see discussion in
Section~\ref{sub:dusty}). A comparison with the LMC CMDs reveals that
sgB[e] stars are on average less luminous; while the RSGs are less dusty
(see Section~\ref{sec:rsg_mdot}).

\begin{figure}[ht]  
\includegraphics[width=3.5in]{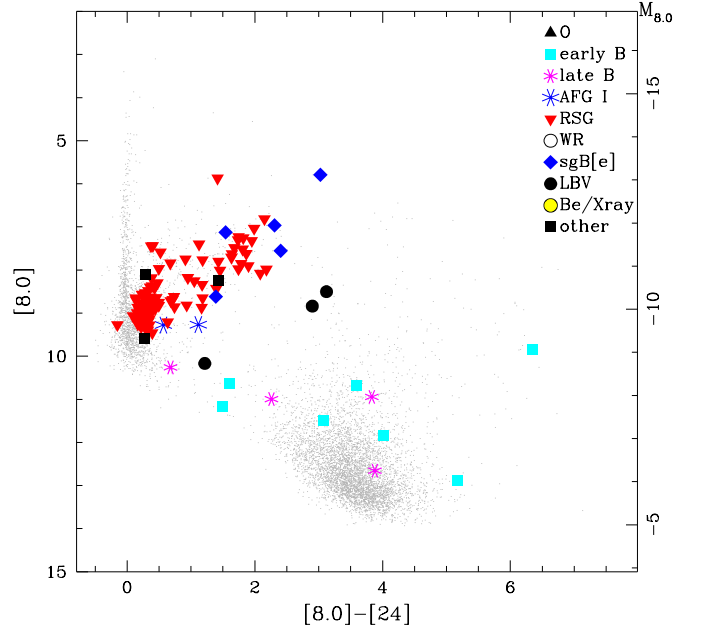}
\caption{[8.0] versus $[8.0]-[24]$ color magnitude diagram, with SAGE
  detections shown as gray points. The RSGs, sgB[e] and LBVs are also
  among the most luminous stars at 8~$\mu$m. The names and
  classifications of stars labeled as ``other'' are (in order of
  increasing [8.0] magnitude): 2dFS0712 (G8), 2dFS3528 (F0), and
  2dFS1829 (G2).}
\label{cmd8}
\end{figure}

\begin{figure}[ht]  
\includegraphics[width=3.5in]{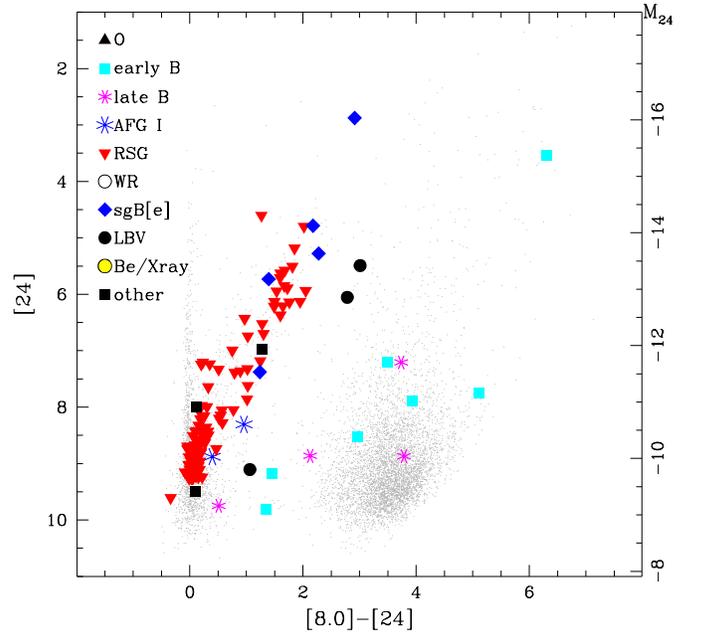}
\caption{Same as Figure~\ref{cmd8}, but for the [24] versus $[8.0]-[24]$
  color magnitude diagram. The brightest sgB[e], RSGs and LBVs are among
  the most luminous stars at 24~$\mu$m. The names and classifications of
  stars labeled as ``other'' are (in order of increasing [24]
  magnitude): 2dFS3528 (F0), 2dFS0712 (G8), and 2dFS1829 (G2).}
\label{cmd24}
\end{figure}

In Figures~\ref{cmd8} and \ref{cmd24}, we present $[8.0]$ and $[24]$
versus $[8.0]-[24]$ CMDs, respectively. At these wavelengths, only stars
with strong [24] emission are detected, as the sensitivity of {\it
  Spitzer} drops sharply, while the flux from hot stellar photospheres
also decreases. Stars without cool dust are located at
$[8.0]-[24]\sim0$, while dusty stars (including RSGs, LBVs, sgB[e]
stars) are reddened. The faint population of sources with
$[8.0]-[24]\sim3.5$~mag corresponds to background galaxies. We detect
dusty LBVs, sgB[e], RSGs, yellow supergiants and some dusty early-B
stars in these CMDs. The detection of the following stars -- labeled as
``other'' -- at [24] (in order of increasing magnitude at [24]):
2dFS3528\footnote{2dFS3528 is discussed further in
  Section~\ref{sub:lbv}.} (F0), 2dFS0712 (G8), and 2dFS1829 (G2),
provides further evidence for their supergiant luminosity class and thus
membership to the SMC. The early B-type stars appearing in these
diagrams, in order of increasing magnitude at [24] are as follows:
2dFS0402, 2dFS2673 (both of type B0-5V; discussed above), AzV\,9
(B0III), AzV\,201 (B1(Be-Fe)), AzV\,409 (or 2dFS2201; B0.5Ib), AzV\,230
(B0Ib) and AzV\,390 (B2Iab:). Of these, both AzV\,409 and AzV\,230 have
spectral energy distributions (SEDs) with disk-like characteristics in
all $Spitzer$ passbands (i.e.\ similar to the Be star SEDs shown in
Paper I, Figure~17), although the former has a slight upturn at [24].
The nature of these objects is discussed in
Section~\ref{sub:dusty}. Finally, the late-B and AFG supergiants present
in these diagrams are hypergiants whose spectra and SEDs indicate the
presence of substantial stellar winds. A comparison of the colors of
various types of massive stars in the LMC versus SMC CMDs reveals that
the colors of the RSGs differ the most (see Section~\ref{sec:rsg}).

\begin{figure*}[ht]  
\includegraphics[angle=270,width=7in]{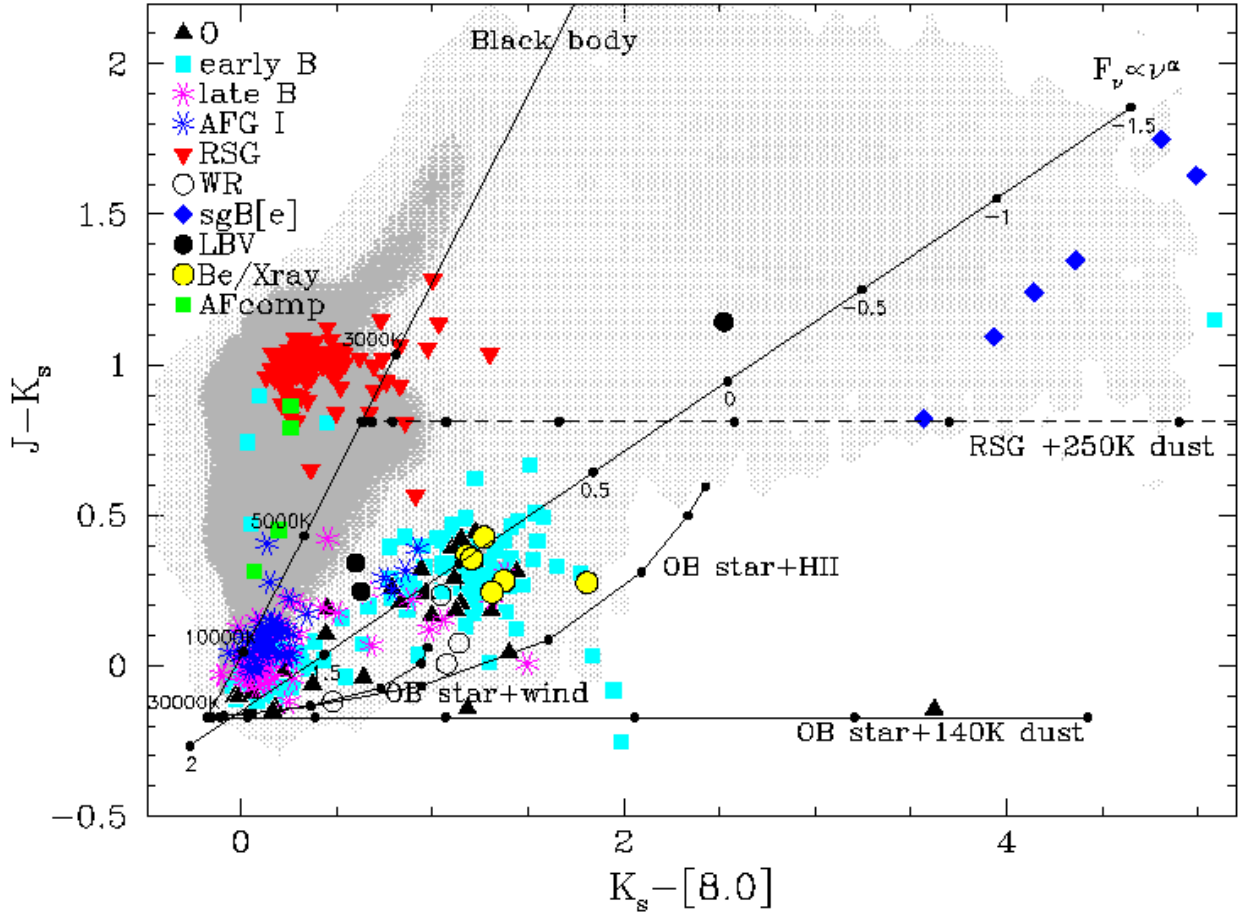}
\caption{$J-K_s$ versus $K_s-[8.0]$ diagram for massive stars in our
  catalog. The locations of all the SAGE detections are shown in gray as
  a Hess diagram. The solid lines represent models (described in Section
  \ref{sec:cmd}): (i) a BB at various temperatures, as labeled, (ii) a
  power law model F$_\nu \propto \nu^{\alpha}$, for
  $-1.5\leq\alpha\leq2$, (iii) an OB star plus an ionized wind, (iv) an
  OB star plus emission from an optically thin \ion{H}{2} region, (v) an
  OB star plus 140~K dust, (vi) 3,500~K blackbody plus 250~K dust
  (dashed line). The sgB[e], RSGs and the clump of ``red'' early-B stars
  that correspond to Be stars occupy distinct regions on this diagram.}
\label{jkk80}
\end{figure*}

\begin{figure*}[ht]  
\includegraphics[angle=270,width=7in]{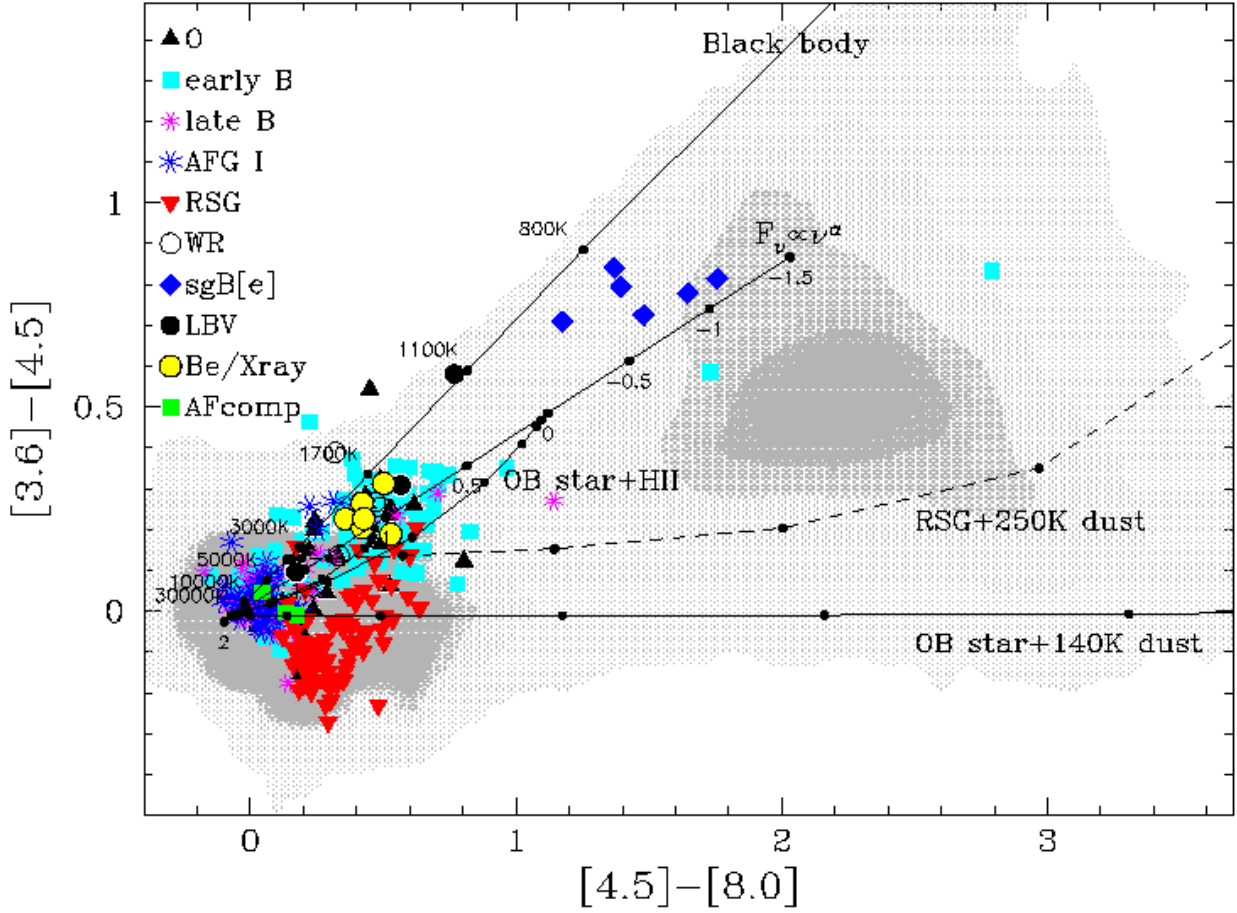}
\caption{Same as Figure~\ref{jkk80}, but for the $[3.6]-[4.5]$
versus $[4.5]-[8.0]$ diagram. The majority of hot massive stars lie between
the blackbody and OB star +wind model, illustrating that a BB is a good
approximation in the infrared.}
\label{iraccc}
\end{figure*}

\begin{figure*}[ht]  
\includegraphics[angle=270,width=7in]{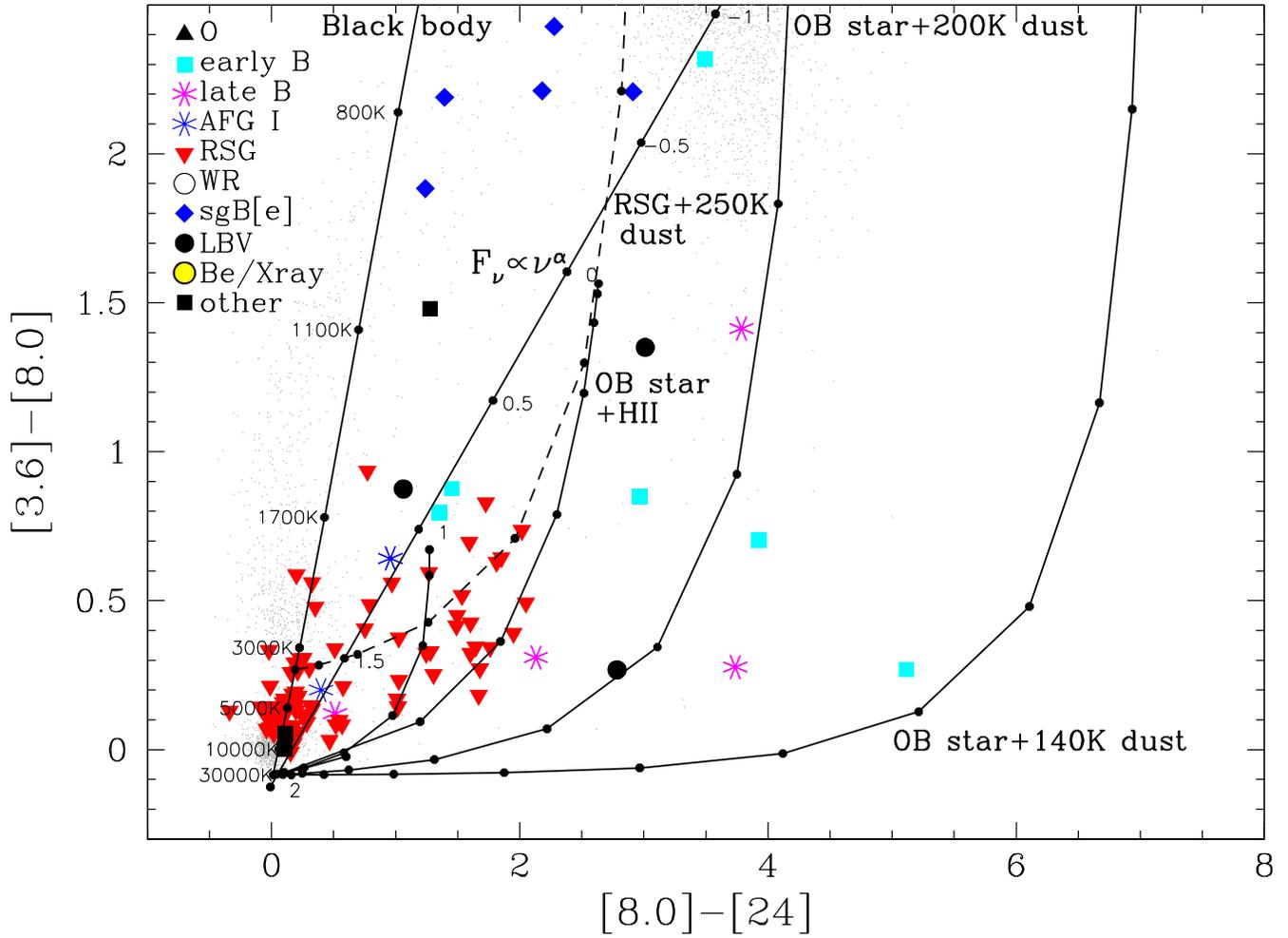}
\caption{Same as Figure~\ref{jkk80}, but for the $[3.6]-[8.0]$
  versus $[8.0]-[24]$ diagram. The more uniform distribution of RSGs, as
  compared to the LMC, is due to the lower dust content.}
\label{iraccc2}
\end{figure*}

In Figures~\ref{jkk80}, \ref{iraccc} and \ref{iraccc2} we present TCDs
using the near and mid-infrared photometry from our matched catalog. We
label stars according to their spectral types and overplot all the
SAGE-SMC detections in gray as Hess diagrams. Stars without dust have
mid-infrared colors near 0. Late-type stars can have $J-K_s$ excesses of
$\leq2$ mag, while the group of stars with colors $K_s-[8.0]\sim3$,
$[4.5]-[8.0]\sim2$ and $[8.0]-[24]\sim3.5$~mag correspond to young
stellar objects and background galaxies. We have overplotted 6 simple
theoretical models to guide the interpretation of the stars in these
diagrams, which are described in detail in Paper I: (i) a blackbody (BB)
at various temperatures, (ii) a power-law model F$_\nu \propto
\nu^{\alpha}$, for $-1.5\leq\alpha\leq2$, (iii) an OB star (represented
by a 30,000~K BB) plus an ionized wind, (iv) an OB star plus emission
from an optically thin \ion{H}{2} region, (v) an OB star plus 140~K
dust, (vi) a 3,500~K BB plus 250~K dust. The most conspicuous stars in
all TCDs are the sgB[e] stars, which have large excesses of $\sim4$~mag
in the $K-[8.0]$ color, similar to the sgB[e] stars in the LMC. R4, a
B0[e]-type LBV, has mid-infrared colors intermediate between sgB[e] and
Be stars. We find a remarkable bimodal distribution of the OB stars,
most clearly seen in Figure~\ref{jkk80}, with the redder grouping
corresponding to Oe and Be stars (see Section \ref{sec:oebe} for more
details). Among these outliers are the Be/X-ray binaries and other
emission line stars, including some yellow supergiants. The
near-infrared excesses found among the sgB[e] and WR stars are on
average lower than in the LMC, presumably due to the lower metallicity
of the SMC. Finally, the RSGs, in contrast to their distribution in the
LMC TCDs, are found to be clustered, due to the lower dust production
rates at the metallicity of the SMC (see Section \ref{sec:rsg_mdot}).

\section{Infrared Excesses of Massive Stars in the SMC versus the LMC}
\label{sec:hotstars}

In this section we study the infrared excesses of specific spectral
types of massive stars, comparing them with the infrared excesses found
in Paper I for the same types of stars in the LMC.

\subsection {O/Oe and early-B/Be stars}
\label{sec:oebe}

\begin{figure*}[ht]  
\includegraphics[angle=270,width=7in]{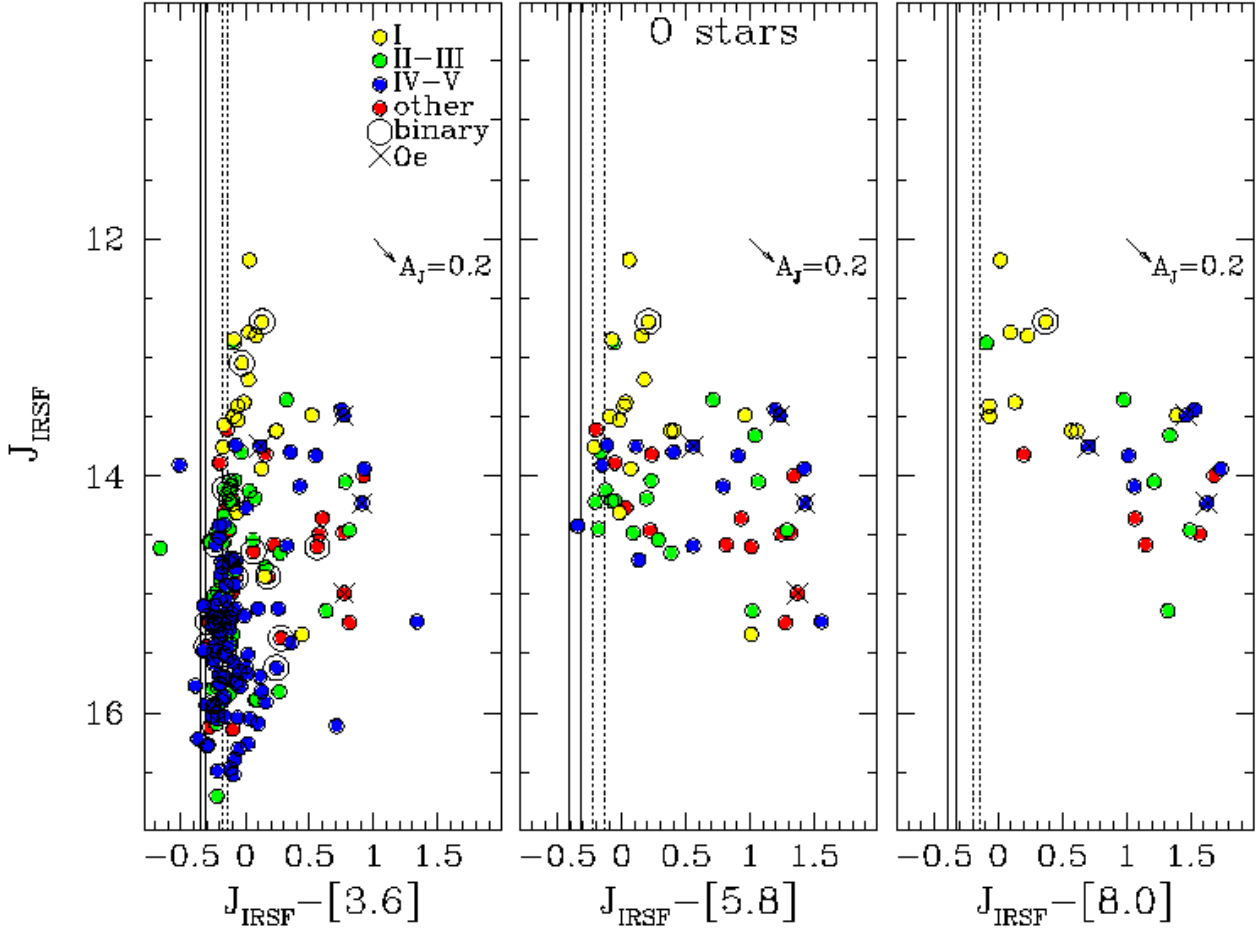}
\caption{Infrared excesses ($J_{IRSF}$ versus $J_{IRSF}-[3.6]$,
  $J_{IRSF}-[5.8]$ and $J_{IRSF}-[8.0]$) for 208 O stars. Supergiants
  are shown in yellow, giants in green, main-sequence stars in blue,
  stars with uncertain classifications (``other'') in red, binaries with
  a large circle and Oe stars with an $\times$. The solid lines
  correspond to 30kK and 50kK TLUSTY models with $\log g = 4.0$. A
  reddening vector for $E(B-V)=0.2$ mag is shown, as well as reddened
  TLUSTY models by this same amount (dotted lines). The more luminous
  stars exhibit larger infrared excesses, which increase with
  $\lambda$.}
\label{fig:oexcess}
\end{figure*}

\begin{figure*}[ht]  
\includegraphics[angle=270,width=7in]{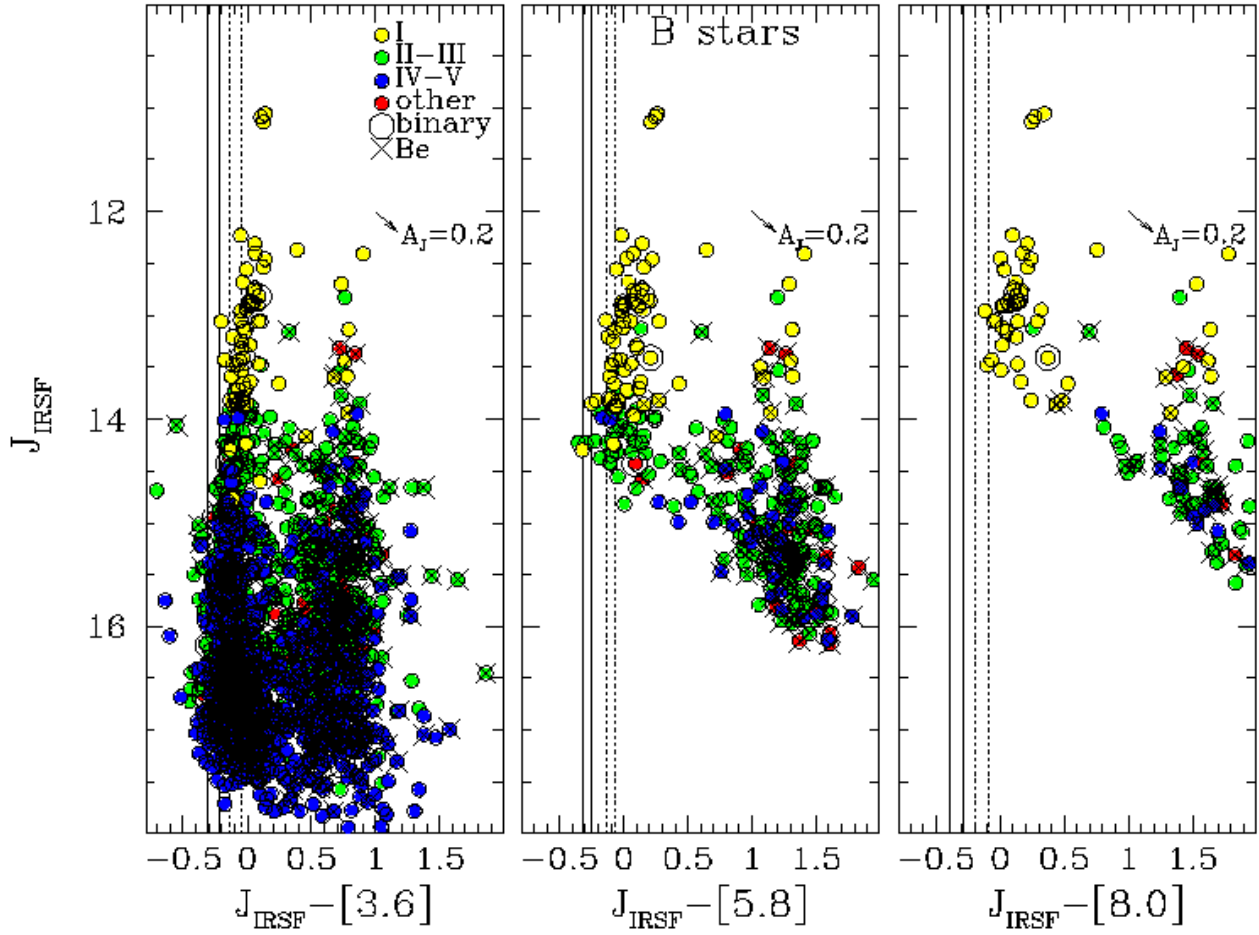}
\caption{Same as Figure~\ref{fig:oexcess}, but for 1967 early-B
  stars. The solid lines correspond to 20kK, $\log g = 3.0$ and 30kK,
  $\log g = 4.0$ TLUSTY models. A reddening vector for $E(B-V)=0.2$ mag
  is shown, as well as reddened TLUSTY models by this same amount
  (dotted lines). The distinct redder sequence, corresponding to the Be
  stars, implies the possibility of identifying or confirming Be stars
  photometrically.}
\label{fig:bexcess}
\end{figure*}

In Figures~\ref{fig:oexcess} and~\ref{fig:bexcess}, we plot $J_{IRSF}$
versus  $J_{IRSF}-[3.6]$, $J_{IRSF}-[5.8]$ and $J_{IRSF}-[8.0]$ colors for
the 208 O and 1967 early-B stars in our catalog, respectively, denoting
their luminosity classes, binarity and emission line classification
properties by different symbols. We compare the observed colors with
colors of plane-parallel non-LTE TLUSTY stellar atmosphere models
\citep{Lanz03, Lanz07} of appropriate metallicity and effective
temperatures. For reference, reddening vectors and TLUSTY models
reddened by $E(B-V)=0.2$ mag are also shown.  We clearly detect infrared
excesses from free-free emission despite not having dereddened the
stars, as in the LMC. At longer wavelengths, the excess is larger
because the flux due to free--free emission for optically thin winds
remains essentially constant with wavelength. Fewer stars are detected
at longer wavelengths because of the decreasing sensitivity of {\it
  Spitzer} and the overall decline of their SEDs. We find that the
majority of early-B supergiants in the SMC exhibit lower infrared
excesses, when compared to their counterparts in the LMC (see Paper I),
due to their lower mass-loss rates, although certain exceptions exist
and deserve further study.

The CMDs allow us to study the frequency of Oe and Be stars, given the
low foreground and internal reddening for the SMC. Our SMC catalog
contains 4 Oe stars among 208 O stars (see Figure~\ref{fig:oexcess}), of
which one is bluer than the rest. There are 16 additional stars with
$J_{IRSF}-[3.6]>0.5$~mag and $J_{IRSF}<15$~mag (including all luminosity
classes), whose spectra appear normal (although the H$\alpha$ spectral
region in most cases was not observed). We refer to these as
``photometric Oe'' stars and attribute their infrared excesses to
free-free emission from a short-lived, possibly recurrent circumstellar
region, whose H$\alpha$ emission line was not detected during the
spectroscopic observations either because the gas had dispersed or
because the region was optically thick to H$\alpha$ radiation or the
observation spectral range just did not extend to H$\alpha$. Given the
expectation of lower $\dot{M}$ at SMC metallicity, we argue that such a
region is more likely to be a transient disk rather than a
wind. Assuming these are all Oe stars, we find a $10\pm2\%$ fraction of
Oe stars among the O stars in the SMC. The error in the fraction is
dominated by small number statistics. In contrast, there are 4 Oe and 14
``photometric Oe'' stars (with $J_{IRSF}-[3.6]>0.5$~mag and
$J_{IRSF}<14.5$~mag) out of 354 O stars in the LMC (despite the higher
$\dot{M}$ at LMC metallicity), which yields a $5\pm1\%$ fraction of Oe
stars among O stars in the LMC.

Turning to the early-B stars, the most striking feature in
Figure~\ref{fig:bexcess} is a distinct sequence displaced by
$\sim0.8$~mag to the red. A large fraction of the stars falling on this
redder sequence have Be star classifications, although not all Be stars
reside there. Given that the circumstellar gas disks responsible for the
emission in Be stars are known to completely vanish and reappear between
spectra taken even 1 year apart \citep[see review by][and references
  therein]{Porter03a}, the double sequence reported here provides
further evidence for the transient nature of the Be phenomenon. A
bimodal distribution at the $L-$band was previously suggested by the
study of \citet{Dougherty94}, which included a sample of 144 Galactic Be
stars. Our larger Be sample, which is essentially unaffected by
reddening, and the inclusion of all early-B stars, clearly confirms the
bimodal distribution. It is due to the much larger number of Be stars
classified in the SMC, in comparison to the LMC, as well as the higher
fraction of Be stars among early-B stars in the SMC, which is $19\pm1\%$
versus $4\pm1\%$ in the LMC when considering only the spectroscopically
confirmed Be stars \citep[cf. $\sim17\%$ for $<10$~Myr B0--5
  stars;][]{Wisniewski06}. Excluding the targeted sample of
\citet{Martayan07a,Martayan07b} does not significantly bias the
statistics, since the fraction only decreases to $15\pm1\%$. We caution
that incompleteness in our catalogs could also affect the determined
fractions, if our sample turns out not to be representative of the whole
population of OB stars.

We proceed to define ``photometric Be'' stars as early-B type stars with
an intrinsic color $J_{IRSF}-[3.6]>0.5$~mag, given that a circumstellar
disk or envelope is required to explain such large excesses. Including
these ``photometric Be'' stars and using the same color and magnitude
cuts as for the ``photometric Oe'' stars above, yields fractions of Be
stars among early-B stars of $27\pm2\%$ for the SMC and $16\pm2\%$ for
the LMC \citep[cf. 32\% from young SMC
  clusters;][]{Wisniewski06}. Table~\ref{tab:oebefractions} summarizes
these results. We compare our results with the fractions determined by
\citet{Maeder99} from young clusters, i.e.\ 39\% for the SMC and 23\%
for the LMC, finding ours to be lower, although the sample selections
were very different.

These preliminary statistics (available for the first time for Oe stars)
indicate that both Oe and Be stars are twice as common in the SMC than
in the LMC. We emphasize the importance of including the ``photometric
Be'' stars, which significantly increase the frequencies of Oe/O and
Be/early-B stars determined and are crucial when comparing such stars in
different galaxies. This novel method of confirming Oe and Be star
candidates from their infrared colors or a combination of their optical
and infrared colors, as recently suggested by \citet{Ita10b} is
complementary to the detailed spectroscopic analyses by
e.g.\ \citet{Negueruela04} on individual Oe stars to understand their
nature, although it is limited to galaxies with low internal reddening.
We finally note that the spectral types of Oe stars in the SMC (O7.5Ve,
O7Ve, O4-7Ve and O9-B0III-Ve) and the LMC (O9Ve (Fe II), O7:Ve,
O8-9IIIne, O3e) are earlier than those of known Galactic Oe stars, which
are all found in the O9-B0 range \citep{Negueruela04}.

\begin{figure}[ht]  
\includegraphics[width=3.5in]{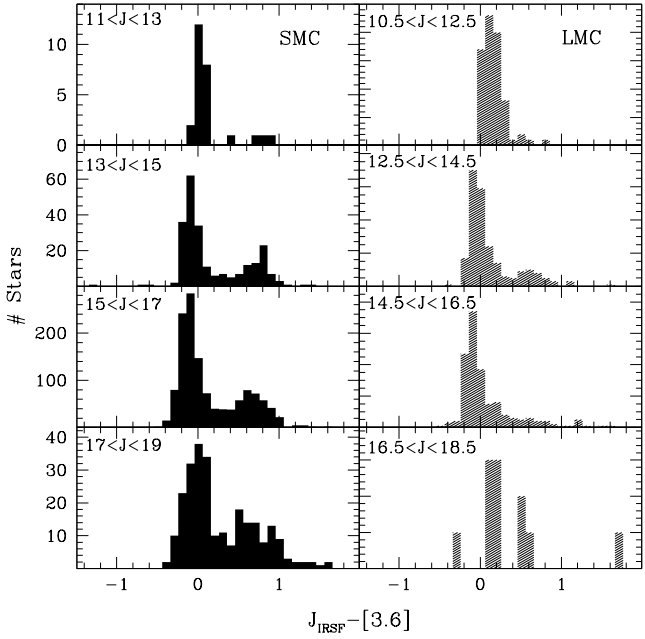}
\caption{Histograms of the J$_{IRSF}-[3.6]$ colors of 1967 early B stars
  in the SMC ({\it left}) and 586 in the LMC (from Paper I, {\it
    right}), divided into magnitude ranges. The main peaks of the
  histograms shift to redder colors for brighter stars. The redder
  secondary peaks correspond to the Be stars, which are more numerous in
  the SMC.}
\label{fig:hist}
\end{figure}

To illustrate the double sequence more clearly, we present the same
information in a histogram in Figure~\ref{fig:hist} for the early-B
stars in both the LMC (from Paper~I) and SMC, divided into magnitude
bins. For stars with $J_{IRSF}-[3.6]\sim0$, we find the brighter, more
luminous stars to be redder (cf.\ top 3 panels of
Figure~\ref{fig:hist}), as in Paper I. In the second and third panels,
the mean values of the $J_{IRSF}-[3.6]$ colors for the (blue, red) peaks
are: ($-$0.07, 0.74) and ($-$0.09, 0.71) mag for the SMC and ($-$0.01,
0.66) and ($-$0.06, 0.79) mag for the LMC. The blue peak for the SMC
stars is thus bluer than for the LMC, which is primarily due to the
higher foreground reddening toward the LMC. In the second panel we find
the separation between peaks to be larger for the SMC (0.81 mag) than
the LMC (0.67 mag), implying that the SMC Be stars have larger infrared
excesses than their LMC counterparts. Unfortunately, in the third panel,
the red peak of the LMC stars is barely discernible, therefore we cannot
confirm this difference. We also ran a Kolmogorov-Smirnov test to check
whether the stars are drawn from the same population (assuming the
sample is unbiased). This yielded a probability of 5.0$\times 10^{-3}$
(for the stars in the 2nd row) and 5.7$\times 10^{-5}$ (3rd row),
implying that the underlying populations are indeed different.

Finally, we note that the brightest Be stars in the SMC
($J_{IRSF}\sim13.2$~mag) are brighter than the brightest Be stars in the
LMC ($J_{IRSF}\sim13.4$~mag), i.e.\ there is a 0.7 mag difference in
absolute magnitude, given the 0.5 mag difference in the distance
moduli. \citet{Garmany85} were the first to make such a comparison,
finding the Be stars in the Magellanic Clouds to be up to 1 mag brighter
when compared to Galactic Be stars, although their observation could be
due to a selection effect, related to the selection of field versus
cluster Be stars. Even though our sample is not dominated by such
selection effects, we also find the brightest Be stars in the SMC to be
intrinsically more luminous than their LMC counterparts. Given that
winds are weaker at low metallicity, it is possible that there is a
mechanism that allows disks to form at higher luminosities\footnote{This
  would also explain the larger infrared excesses found in Be stars in
  the SMC, described in the previous paragraph.}. Such a possibility
would have interesting implications when extrapolating to even lower
metallicities \citep[see, e.g.][]{Ekstrom08}.

\subsection{Be/X-ray Binaries}

\citet{Dray06} found that both metallicity and a star formation burst
must be invoked to explain the large number of high mass X-ray binaries
in the SMC, and consequently the large number of Be/X-ray binaries. Of
the 69 Be/X-ray binaries in our SMC photometric catalog, 21 (or
$30\pm8\%$) were matched in the SAGE-SMC database (versus 4 matches out
of 20 Be/X-ray binaries or $20\pm11\%$ in the LMC). Interestingly, some
SMC Be/X-ray binaries are brighter than their LMC counterparts, despite
being farther away, in agreement with the finding that Be stars in the
SMC occur at higher luminosity. A comparison with the infrared colors of
the Be/X-ray binaries in the LMC does not reveal any differences. The
SEDs of the SMC Be/X-ray binaries show excess due to the disks around
the Be stars, as in the LMC. Finally, the multiple $VI$ and $JHK_s$
measurements reveal variability for many systems, which deserves further
study.

\subsection{Wolf-Rayet stars}
\label{sub:wr}

Of the 12 WR stars in our spectral-type catalog, 10 were matched in the
SAGE-SMC database. HD\,5980 (SMC-WR5;AzV\,229 WN6h;LBV binary) is
labelled as an LBV in all diagrams, and is the only WR star with a
detection at [24]. We compare the colors of the WR stars to those of
their LMC counterparts (see Paper I) and find that their infrared
excesses are much lower, possibly indicating that their winds are much
weaker than those of LMC WR stars. Specifically, we find the SMC WR
stars to have ``bluer'' colors: $-0.2<J-[3.6]<0.5$ (versus $0.2-1.7$ for
the LMC sample), $-0.2<J-K_s<0.1$ (versus $0-0.6$),
$0.4<K_s-[8.0]<1.2$~mag (versus $0.4-1.8$). The origin of the hydrogen
absorption lines seen in 10 of the 12 WR stars (classified as ``WN+abs''
or ``WNha'') remains unclear, since radial velocity surveys have only
confirmed 4 as binary systems \citep{Foellmi03}. The absorption in the
other stars could alternatively be explained as photospheric absorption
\citep[due to weak winds;][]{Massey03c}. The infrared photometry made
available for the WR stars in this paper, will enable modeling of their
SEDs to determine the properties of their stellar winds and perhaps the
origin of the absorption lines.

\subsection{AF composite stars}
\label{sub:afcomp}

\citet{Evans04} discovered a class of objects with peculiar composite
spectra (AF composites or AFcomp), which yield an A-type classification
from their \ion{Ca}{2}~K lines, but an F-type classification from other
metal lines and the G-band. For example the classification AFA3kF8
denotes such an AF composite star, with \ion{Ca}{2}~K like A3 and G-band
like F8. Our photometric catalog contains 16 out of the 20 AF composite
stars from the 2dF Survey of \citet{Evans04}. The AF composite stars
that additionally exhibit forbidden emission lines
(2dFS1804;AFA3kF0/B[e] and 2dFS2837;AFA5kF0/B[e]) are labeled and
henceforth treated as sgB[e] stars.

The infrared colors of the AF composite stars provide further clues to
their nature and confirm their astrophysical peculiarity. In
Figure~\ref{cmd36}, their $[3.6]-[4.5]$ colors are indistinguishable
from AFG supergiants\footnote{The brightest AF composite star at [3.6]
  is 2dFS5049.}, while in Figure~\ref{cmd36j36} they clearly have red
colors, with $0.5<J-[3.6]<1$~mag. None are detected at [24]. In
Figure~\ref{jkk80}, all 4 AF composite stars with [8.0] detections show
near-infrared colors in the range $0.3<J-K_s<0.9$~mag; thus,
intermediate between blue and red supergiants, possibly indicating these
stars are on blue loops or in interacting binaries. The SEDs indeed are
quite ``flat''; thus, not consistent with normal single stars, while the
multiple $VI$ and $JHK_s$ measurements reveal variability for several
sources, which in some cases can be confirmed by light curves available
from the OGLE database \citep{Udalski97,Szymanski05}. We note that
2dFS2945 (HV 11519) is a known $\delta$ Cep pulsator.

\subsection{Dusty stars}
\label{sub:dusty}

\begin{figure}[ht]
\includegraphics[width=3.5in]{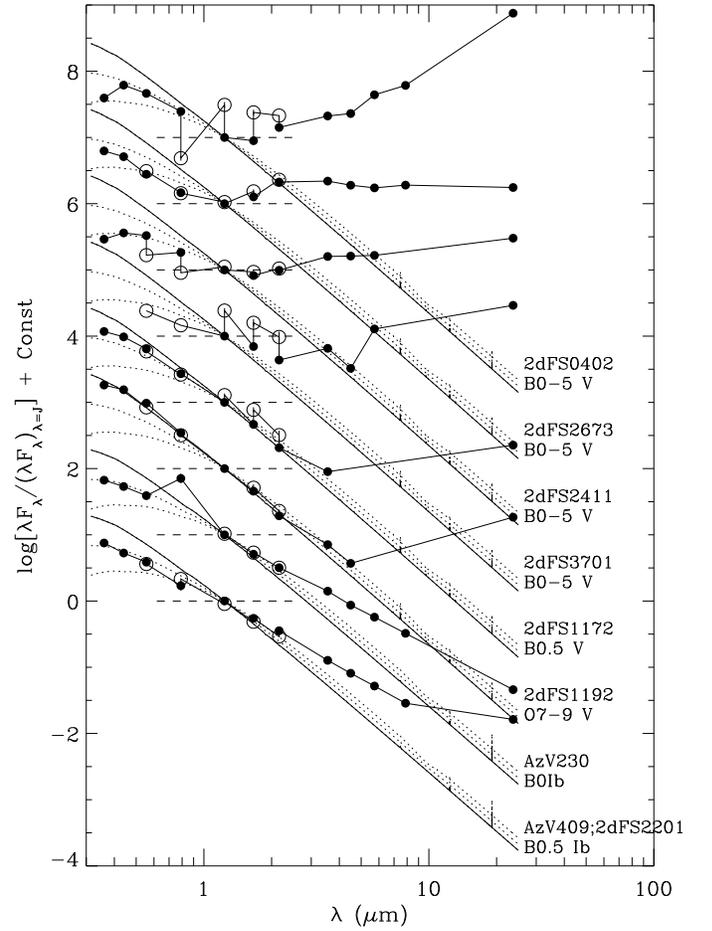}
\caption{SEDs of OB stars with [24] detections, normalized by their
  $J-$band fluxes (dashed line) and offset for display
  purposes. Normalized TLUSTY model atmospheres at SMC metallicity
  (30kK, $\log g = 3.00$ for the main sequence stars; 20kK, $\log g =
  2.25$ for the 2 supergiants) are overplotted as solid lines for
  comparison. The MCPS, IRSF and SAGE measurements (filled circles) and
  the 2MASS and OGLE measurements (open circles) are also connected by a
  solid line. The dotted curves correspond to TLUSTY models reddened by
  $E(B-V) = 0.25$ and 0.50 mag. Several classes of object are
  represented (see Section~\ref{sub:dusty}).  2dFS1172 and 2dFS1192 are
  examples of dusty OB stars. Variability is present in several sources;
  however, in the case of 2dFS0402 it is likely due to a mismatch with a
  close infrared source.}
\label{fig:pecstack}
\end{figure}

\begin{figure}[ht]
\includegraphics[width=3.5in]{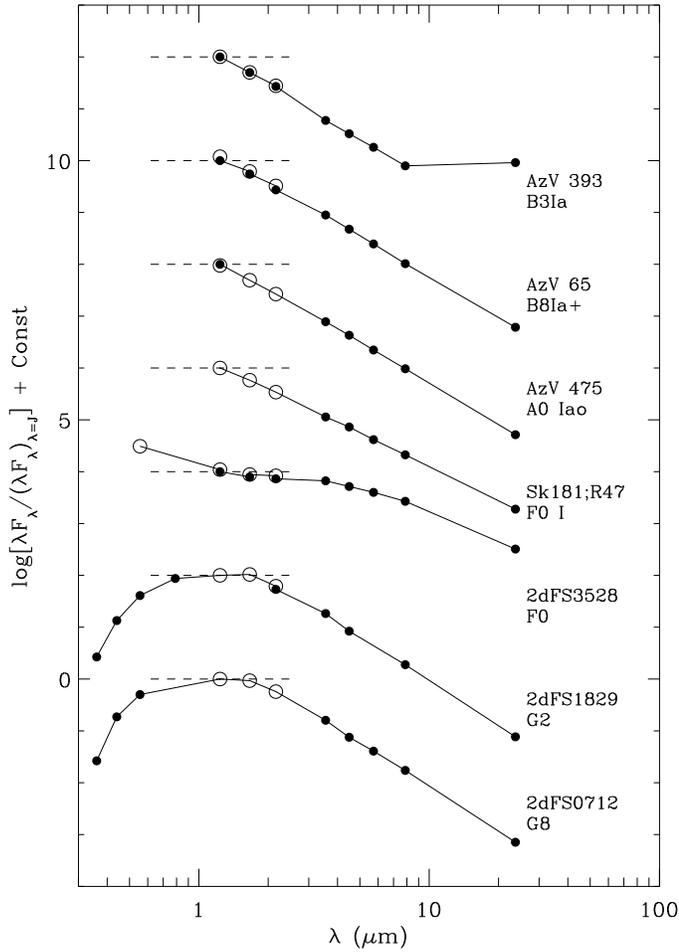}
\caption{SEDs of 2 late-B supergiants and of the 5 AFG supergiants
  detected at [24], normalized by their $J-$band fluxes (dashed line)
  and offset for display purposes, as in Figure~\ref{fig:pecstack}. The
  SED of the F~supergiant 2dFS3528 most resembles that of R4 (AzV\,16,
  B0[e]LBV); we suggest it to be an LBV candidate.}
\label{fig:afgstack}
\end{figure}

Our catalog includes 166 stars with detections at [24], which indicate
the presence of dust. Of these, 97 are RSGs, 5 sgB[e] and 3 LBV stars,
all exhibiting dust, as expected. Luminous supergiants with strong winds
or disk-like SEDs, also exhibit infrared excess at [24] and [8.0] and
therefore appear in Figures~\ref{cmd8} and~\ref{cmd24}. There are 4 B0-5
V type stars: 2dFS0402, 2dFS2411, 2dFS2673, and 2dFS3701 that have very
flat SEDs. Additionally, we find a class of 44 dusty OB stars (16 late-O
and 28 early-B stars, making up 2\% of our sample), with similar
characteristics to the peculiar population of dusty OB stars reported by
\citet[][190 O9-B3 stars, corresponding to 5\% of their
  sample]{Bolatto07} in their S$^{3}$MC survey and recently confirmed by
\citet{Ita10a} with data from AKARI.  Hereafter, we refer to these stars
as ``dusty OB stars''. They are characterized by a photospheric SED out
to [5.8] and strong [24] excess, which cannot be attributed to free-free
emission. Most have main-sequence luminosity classes, although there are
a few giants and supergiants. In Figure~\ref{fig:pecstack}, we plot SEDs
for the 4 stars with unusually flat or rising SEDs, 2 representative
examples from among the 44 ``dusty OB stars'' (2dFS1172, 2dFS1192) and 2
examples of luminous supergiants with infrared excess due to strong
winds or disks (AzV\,230, AzV\,409;2dFS2201). For completeness, in
Figure~\ref{fig:afgstack}, we show 2 examples of late-B supergiants with
[24] detections and the 5 AFG supergiants that have [24]
detections. Note, the F0 supergiants R47 \citep[a rare eclipsing
  binary;][]{Prieto08c} and 2dFS3528 (an LBV candidate, see
Section~\ref{sub:lbv}) are of particular interest.

The dust detected in the class of ``dusty OB stars'' could either be
directly related to the OB stars or simply be contamination given the
volume of the SMC sampled with the $6\arcsec$ resolution of MIPS.  While
inspection of their environments shows that they are located in or near
young associations, they are frequently not the youngest objects
(earliest types), e.g.\ NGC\,330 has 3 ``dusty OB stars'' nearby and a
turn-off of around B2 \citep{Lennon93}. We examined the spectra for 18
of these sources \citep[12 from the 2dF Survey and 6 from the VLT-FLAMES
  Survey;][]{Evans06}, which all exhibit sharp nebular-like Balmer
emission lines; for those with red-optical spectra, 7 also have
H$\alpha$ nebular emission and 3 have weak [\ion{N}{2}] and [\ion{S}{2}]
lines, indicating the presence of a faint \ion{H}{2} region. Many
sources for which we do not have spectra available are either listed in
H$\alpha$ emission line catalogs or lie in regions of faint nebulosity.
Given that most ``dusty OB stars'' are not found within very young
regions, we suggest that these sources represent unrelated cirrus hot
spots or cases where hot stars are ionizing nearby dusty molecular
clouds (rather than young stellar objects) producing both the nebular
emission lines and emission from warm dust \citep[as in Cepheus B, see
  e.g.][]{Panagia81}.

\subsubsection{Supergiant B[e] stars}
\label{sub:sgbe}

In the SMC photometric catalog, we have detected 7 luminous sources with
colors typical of sgB[e] stars (see Paper I for an introduction),
i.e.\ $M_{3.6}<-8$, $[3.6]-[4.5]>0.7$, and $J-[3.6]>2$~mag (see
Figures~\ref{cmd36} and~\ref{cmd36j36}). Five of these are previously
known sgB[e] stars (with R50; B2-3[e] being the brightest in all IRAC
and MIPS bands), while R4 (AzV\,16) is classified as an LBV with a
sgB[e] spectral type. In addition to these, we find that 2dFS1804
(AFA3kF0/B[e]) has a very similar SED (and therefore infrared colors) to
the known sgB[e] 2dFS2837 (AFA5kF0/B[e]).  \citet{Evans04} also remarked
on the similarity of their spectra. We therefore confirm the supergiant
nature of 2dFS1804 and present its SED in Figure~\ref{fig:sgbestack}
along with those of other sgB[e] and LBV stars. We note that both are
photometrically variable. The similarity of the SEDs of these sgB[e]
stars, despite the various optical spectral classifications, implies
that all are the same class of object. The cooler, composite spectral
types indicate a lower mass and perhaps a transitional stage to or from
the sgB[e] phenomenon. The only difference we find between the sgB[e]
stars in the SMC versus the LMC is that on average they are $\sim$1-2 mag
fainter (in absolute terms).

\subsubsection{Luminous Blue Variables}
\label{sub:lbv}

\begin{figure}[ht]
\includegraphics[width=3.5in]{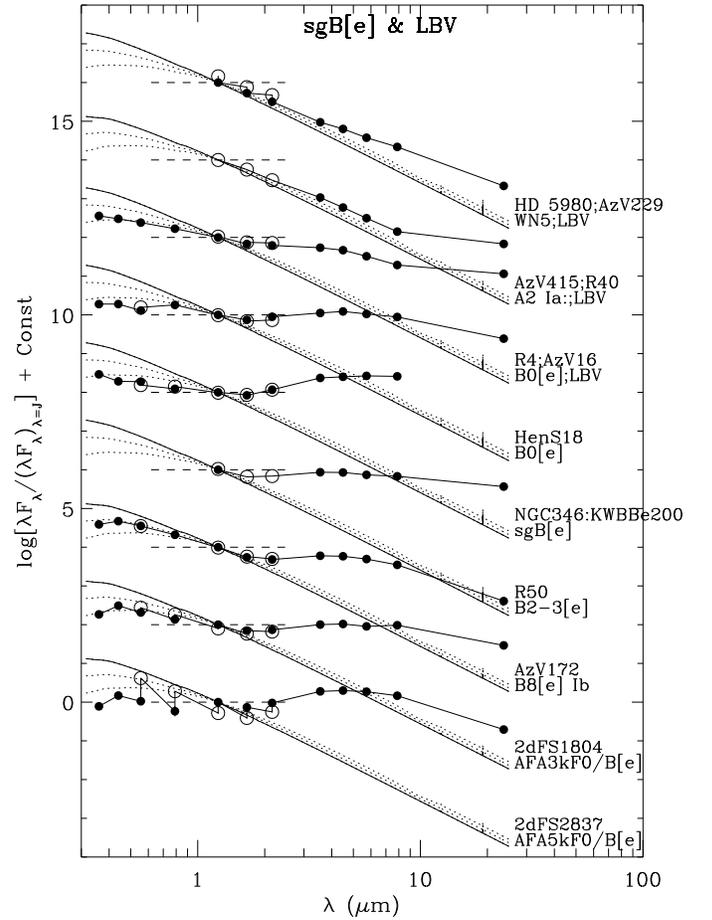}
\caption{SEDs of the LBVs and sgB[e], all detected at [24], normalized
  by their $J-$band fluxes (dashed line) and offset for display
  purposes, as in Figure~\ref{fig:pecstack}. TLUSTY model atmospheres
  are overplotted for comparison: 15kK, $\log g = 1.75$ for the B8 and
  later types; 20kK, $\log g = 2.25$ for the earlier types. The last two
  SEDs with composite spectral types closely resemble the sgB[e] SEDs;
  note, these stars vary.}
\label{fig:sgbestack}
\end{figure}

All 3 known LBVs in the SMC: R4 (AzV\,16, B0[e]LBV), R40 (AzV\,415,
A2Ia: LBV) and HD\,5980 (WN6h;LBV binary), were all detected at infrared
wavelengths. In all CMDs and TCDs, R4 is the more reddened LBV. It
occupies a location similar to other sgB[e] stars on the CMDs; however,
in Figures~\ref{jkk80} and~\ref{iraccc} it lies between the sgB[e] and
the Be stars, whereas the colors of HD\,5980 \citep[a well known
  eccentric eclipsing binary, see e.g.][]{Foellmi08} are similar to
those of the LBVs in the LMC. We plot their SEDs in
Figure~\ref{fig:sgbestack} and find them to differ, given their very
different spectral types. Moreover, we find evidence for variability,
which can be confirmed from existing light curves in the All Sky
Automated Survey \citep[ASAS;][]{Pojmanski02}, as pointed out by
\citet{Szczygiel10}, who studied the variability of the massive stars
presented in Paper~I in the LMC. The various SED shapes and spectral
types observed depend on the time since the last outburst event and the
amount of dust formed.

Finally, we note that the F0 star, 2dFS3528 (Sk166) also lies among the
sgB[e] stars in Figures~\ref{cmd8}, \ref{cmd24} and~\ref{iraccc2} and
near R4 (AzV\,16; B0[e]LBV) in the other CMDs and TCDs, because it has a
similar SED (cf.\ \ref{fig:afgstack}
and~Figures~\ref{fig:sgbestack}). Although Sanduleak originally
classified this star as ``OB:'', its 2dF spectrum is clearly F-type, but
with strong H$\alpha$ emission having broad wings, which indicate a
wind. \citet{Osmer73} designated it as a nonmember based on its optical
colors. It is included in the catalog of
\citet{Massey02}\footnote{[M2002] 80339; $V=12.92$, $B-V=0.15$,
  $U-B=-0.45$, and $V-R=0.17$ mag.} and indeed has some colors that are
too blue, while others are too red for its spectral type.  However, the
radial velocity measured by \citet[][$+137\pm8$~km~s$^{-1}$]{Evans08}
firmly places it in the SMC. The SED of 2dFS3528 is distinctly
sgB[e]-like and most similar to that of R4. All the above
characteristics, along with the variability detected in the
near-infrared, make this star an LBV candidate. Further monitoring is
highly desirable to confirm its LBV nature.

\subsubsection{Red Supergiants}
\label{sec:rsg}

The RSGs are among the brightest mid-infrared sources in the SMC, with
absolute magnitudes reaching $-12$ mag at [3.6], and $-13$ mag at [8.0]
and [24]. However, in the LMC, some RSGs occur at higher [24]
luminosities. SMC\_018592 (K0-2I) is the brightest RSG at [3.6], [8.0]
and [24] (see Figures~\ref{cmd36}, \ref{cmd8}, \ref{cmd24}). In
Figure~\ref{cmd36j36}, the following 3 RSGs have ``bluer'' colors than
the rest (in order of increasing [3.6] magnitude): SMC\_030135 (K2I),
SMC\_035231 (K2I) and SMC\_049990;2dFS1517. The latter has two
classifications: K5I by \citet{Massey03b} and G2 by \citet{Evans04},
making it a variable candidate \citep[given also its radial velocity:
  $+143\pm5$~km~s$^{-1}$,][]{Evans08}. In Figure~\ref{jkk80},
SMC\_035231 (K2I) is the outlier at (0.4, 0.6), making it a possible
foreground dwarf; while SMC\_058738 (K2I) is the outlier at (1.0,
0.5). We note that SMC\_055188 (M3-M4I) and SMC\_046662 (M0I) were
reported by \citet{Levesque07} to have variable spectral types; they are
included in our catalog and are among the more reddened RSGs.

A comparison of the SMC CMDs and TCDs to the corresponding ones of the
LMC from Paper I, reveals that the distribution of RSG colors is quite
different. The SMC RSGs have a dominant population of non-dusty RSGs and
a population of dusty RSGs, with a distinct color distribution. In
contrast, almost all RSGs in the LMC are dusty to some degree, with a
rather continuous distribution in color. This difference is due to the
lower dust content ($\propto Z$), which is not compensated for by the
lower wind velocity \citep[$\propto \sqrt Z$, see][]{vanLoon00,
  vanLoon06, vanLoon08}. Furthermore, since both the dust production
rate \citep[cf.][]{Sloan08} and the dust-to-gas ratio depend on $Z$
\citep{vanLoon06}, $\dot{M}$ is independent of $Z$, though the wind
velocities of RSGs depend on metallicity \citep[see e.g.][]{Habing94,
  Marshall04}. While the optical and near-infrared colors of RSGs have
long been known to display a $Z$ dependence
\citep[cf.][]{Arp59,Oestreicher97}, the systematics of their
mid-infrared colors have only been investigated for limited samples
\citep[e.g.][]{Groenewegen09}.

\section{Red Supergiant Mass-Loss Rates}
\label{sec:rsg_mdot}

\begin{figure}
\includegraphics[width=3.5in]{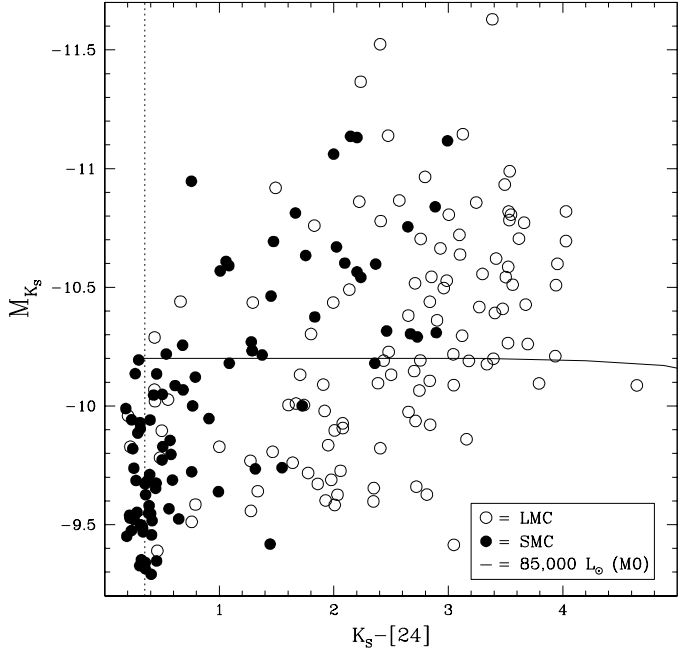}
\caption{$M_{\rm K_s}$ versus $K_s-[24]$ CMD of our sample of RSGs in
  the SMC (filled circles; this work) and LMC (open circles; Paper
  I). The fainter SMC RSGs cluster around $K_s-[24]=0.35$ (dotted
  line). A model from \citet{Groenewegen06} for a star with spectral
  type M0 and luminosity 85,000 L$_\odot$ with increasing amounts of
  dust toward redder color is shown for reference. Most LMC stars and
  many SMC stars show clear excess emission due to dust, which is most
  pronounced at high luminosity.}
\label{rsg}
\end{figure}

Our large samples of RSGs in the SMC and LMC lend themselves to a study
of the $Z$ dependence of the mass-loss rate. Figure~\ref{rsg} presents
the $M_{K_s}$ versus $K_s-[24]$ CMD for both SMC and LMC RSGs (from Paper
I), on which the RSGs are spread out as a result of various amounts of
circumstellar dust. The fainter RSGs in the SMC concentrate around a
vertical sequence at $K_s-[24]\approx0.35$ mag, but all other RSGs have
significantly redder $K_s-[24]$ colors indicative of the emission from
dust. A series of models for a star with spectral type M0 and luminosity
85,000 L$_\odot$ \citep{Groenewegen06} are plotted for reference,
showing that the dust remains optically thin at near-infrared
wavelengths for $K_s-[24]<5$ mag.

\begin{figure}
\includegraphics[width=3.5in]{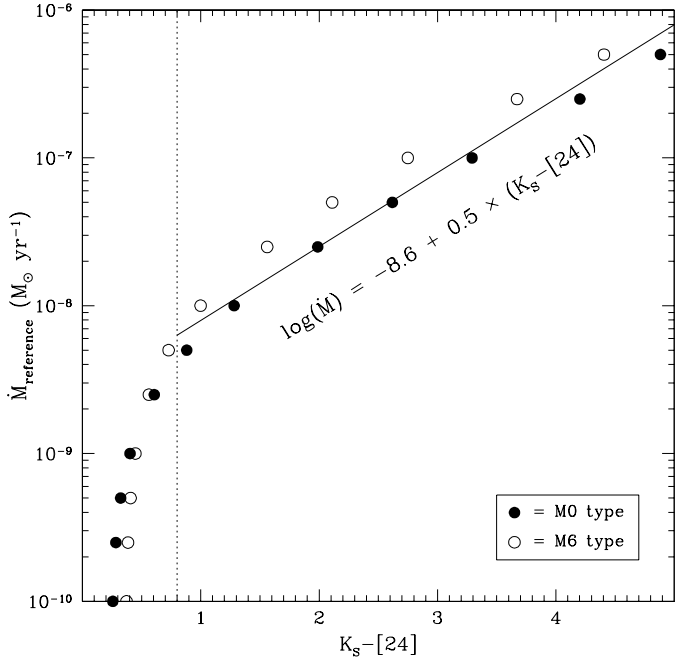}
\caption{$\dot{M}$ as a function of $K_s-[24]$ for two sets of models
  from \citet{Groenewegen06} with spectral types M0 (filled circles) and
  M6 (open circles). The solid line indicates the approximation
  described by the labeled equation, which is valid for a star of
  luminosity 3,000 L$_\odot$, wind speed 10 km s$^{-1}$ and dust-to-gas
  ratio 0.005, for the range $0.8 < K_s-[24] < 5$ mag.  We use scaling
  relations to apply it to the RSGs in the Magellanic Clouds.}
\label{mdot}
\end{figure}

The $K_s-[24]$ color is a good measure of the optical depth of the
circumstellar dust envelope. We use the results from radiative transfer
modeling presented in \citet{Groenewegen06} for a star of luminosity
$L=3000$ L$_\odot$, wind speed $v=10$ km s$^{-1}$, dust-to-gas ratio
$\psi=0.005$, surrounded by silicate grains with a size of 0.1 $\mu$m
and a condensation temperature of 1000~K. The relation between this
reference $\dot{M}$ and $K_s-[24]$ color is shown in Figure~\ref{mdot},
for a star of spectral type M0 (typical for the SMC stars) and again for
a star of spectral type M6 (at the cool extreme of the LMC stars in our
sample). For colors $K_s-[24]<0.8$ mag, the relation is rather steep;
given the inaccuracies in photometry and stellar parameters we will not
assign mass-loss rates to these non-dusty stars. For colors
$0.8<K_s-[24]<5$ mag, the relation is quite well approximated by the
formula:
\begin{equation}
\log(\dot{M}_{\rm reference})=-8.6+0.5\times(K_s-[24]).
\end{equation}
The deviations from this formula are not (much) more than a factor of
two for individual stars, and will tend to average out when considering
the sample as a whole.

The true $\dot{M}$ is obtained by applying scaling relations valid for
dust-driven winds \citep[cf.][]{vanLoon07}:
\begin{equation}
\dot{M}=\dot{M}_{\rm reference}(v/10)(0.005/\psi)(L/3000)^{0.5},
\end{equation}
where the wind speed scales as \citep[e.g.][]{Marshall04}
\begin{equation}
v=10(\psi/0.005)^{0.5}(L/3000)^{0.25}.
\end{equation}
For the dust-to-gas ratio $\psi$, we adopt a value roughly in proportion
to the metallicity of the stellar population, namely, $\psi\approx0.001$
for the SMC and $\psi\approx0.002$ for the LMC \citep[cf.][]{vanLoon00,
  vanLoon06}. The luminosity is obtained by scaling the $K_s-$band
magnitude of the model star from \citet{Groenewegen06}; for the SMC
stars we adopt $K_s=8.1$ mag for the 3000 L$_\odot$ star, at a reference
distance of 8.5 kpc (this corresponds to a spectral type M0) and
$K_s=7.9$ mag for the LMC stars (spectral type about M2). The distance
modulus to the SMC is taken as 18.91 mag, and to the LMC as 18.41 mag
(to be consistent with Paper I). The inaccuracies in the luminosities
resulting from this approximation are of the order of 20\%, which is
small for our purposes.

\begin{figure}
\includegraphics[width=3.5in]{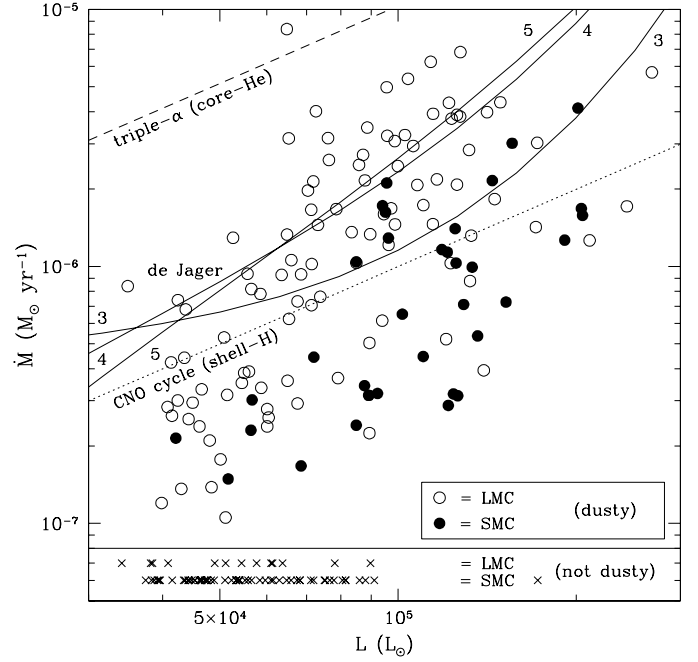}
\caption{$\dot{M}$ versus luminosity L for our samples of RSGs in the
  SMC (filled circles; this work) and LMC (open circles; Paper I). No
  mass-loss rates could be determined for stars with $K_s-[24]<0.8$ mag;
  these are shown in the bottom panel at their respective luminosities
  (crosses). Shown for reference are the nuclear burning rates for the
  triple-$\alpha$ process at work in helium burning in the cores of RSGs
  (dashed line), as well as the CNO cycle in hydrogen shell-burning
  (dotted line) and the prescription of \citet{deJager88} for 3000, 4000
  and 5000K (solid lines, labeled by the first digit of the
  temperature).}
\label{rsg_mdot}
\end{figure}

The estimated mass-loss rates show a clear positive correlation with
luminosity (see Figure~\ref{rsg_mdot}), confirming previous results
\citep{vanLoon99, vanLoon05, vanLoon06, Groenewegen09} with a larger
sample of stars. Apparently, the fainter RSGs are not very proficient in
their mass loss. However, these mass-loss rates were derived by assuming
that dust condenses in these winds at the efficiencies expected from
their metal content. This may not be true, particularly in the
relatively warm atmospheres of early-M or late-K type stars, where the
dust-to-gas ratio can be much less than assumed here
\citep[cf.][]{vanLoon05}. Indeed, at $L<9\times10^4$ L$_\odot$ most of
the RSGs in the SMC show very little evidence of dust at all, but they
may still exhibit a purely gaseous wind. On the other hand, the values
for the wind speed are estimated to be 9--12 km s$^{-1}$ for the SMC
sample (12--19 km s$^{-1}$ for the LMC sample). This is consistent with
expectations for a fluid in which the grains are well coupled to the gas
and thus can transfer momentum from the radiation field onto the bulk
matter \citep[cf.][]{Marshall04}; much lower values could bring this
assumption into question \citep[cf.][]{McDonald09}.

The mass-loss rate is next compared to the rate at which matter is
converted into energy in the central engine of the star, to assess
whether much mass is lost before the star is taken elsewhere on its
evolutionary path. In the context of core-helium burning RSGs, the most
relevant process is the triple-$\alpha$ process, which has an efficiency
$\eta\approx5.9\times10^{17}$ erg g$^{-1}$ \citep{Kippenhahn90}; we also
compare to the CNO cycle, which governs hydrogen shell burning and has
$\eta\approx6.1\times10^{18}$ erg g$^{-1}$. The nuclear burning rate (in
M$_\odot$ yr$^{-1}$) is then:
\begin{equation}
\dot{M}_{\rm nuclear}=6.087\times10^7(L/\eta).
\end{equation}
We find that essentially all RSGs in our samples lose mass through
stellar winds at a significantly lower rate than that at which helium is
burned in their cores (Figure~\ref{rsg_mdot}), often by an order of
magnitude difference. This is in stark contrast to samples of RSGs that
were selected from among the brightest mid-infrared sources;
\citet{vanLoon99} showed that most of those RSGs have mass-loss rates
similar to the nuclear burning rate, with a few extreme objects
exceeding that rate by a significant amount. The optically selected RSGs
in the SMC and LMC are therefore found to be in a more stable phase of
RSG evolution, in contrast to the RSGs selected as bright mid-infrared
sources, which have a shorter lifetime. The prescriptions of
\citet{deJager88} for 3000, 4000 and 5000K coincide with the most
luminous LMC RSGs, but are too high for the less luminous RSGs (in both
galaxies). The 3000K prescription of \citet{deJager88} better reproduces
the luminous RSGs in the SMC, whereas the warmer prescriptions better
reproduce the luminous RSGs in the LMC. This is in conflict with the
observed differences in temperature, with the SMC RSGs having earlier
spectral types than their LMC counterparts.

The SMC sample on the whole exhibits lower dusty mass-loss rates than
the LMC sample (Figure~\ref{rsg_mdot}), and RSGs in the SMC appear to
contribute relatively more at higher luminosities ($L>9\times10^4$
L$_\odot$), while RSGs in the LMC also contribute noticeably at fainter
luminosities. Both observations are likely related to the warmer
atmospheres (earlier spectral types) of the SMC RSGs compared to those
in the LMC. We thus conjecture that dusty mass loss plays a less
important role in the SMC than in the LMC, while other forms of mass
loss may instead be more important. For instance, as recently suggested
\citep{vanLoon10a}, the warmer winds of metal-poor RSGs could be more
susceptible to the coupling of Alfv\'en waves, which could provide a
driving mechanism in the absence of radiation pressure.

We emphasize that our samples miss the dustiest RSGs, which are in brief
episodes of very high mass loss; therefore, our results only apply to
the optically bright RSG phase. A few such very dusty RSGs \citep[known
  to exist in the Magellanic Clouds, e.g.][]{Wood92, Groenewegen98,
  vanLoon99, vanLoon10a, vanLoon10b, Srinivasan09} contribute more dust
than the entire samples of optically bright RSGs considered here.

\section{Summary}
\label{sec:summary}

This paper presents the first catalogs of accurate spectral types and
multi-wavelength photometry of massive stars in the SMC, which are used
to study their infrared properties. The spectroscopic catalog contains
5324 massive stars, with accurate positions and spectral types compiled
from the literature, and constitutes the largest such catalog currently
available for any galaxy. The photometric catalog comprises uniform
0.3--24~$\mu$m photometry in the $UBVIJHK_s$+IRAC+MIPS24 bands for a
subset of 3654 stars that were matched in the SAGE-SMC database. The low
foreground reddening toward the SMC, and the identical distance of the
stars minimize systematic errors due to reddening and enable the
investigation of infrared excesses. As in Paper I, we construct CMDs and
TCDs, and discuss the position of O, early and late-B stars, WR, LBV,
sgB[e], classical Be stars, RSGs, AFG supergiants, and Be/X-ray binaries
on them. These diagrams are useful for interpreting infrared photometry
of resolved massive stars in nearby galaxies at low metallicity.

A comparison of the infrared colors of the massive stars in the SMC to
those of their counterparts in the LMC (presented in Paper I) reveals
differences that are due to the different evolution at SMC
metallicity. The main results of our study concern the emission line Oe
and Be stars, and the RSGs. We detected a clear bimodal distribution of
early-B stars, with the redder sequence corresponding to classical Be
stars and therefore propose that Be stars (and similarly Oe stars) can
be easily discriminated photometrically, by their infrared colors.  We
find the fraction of emission line stars in the SMC ($10\pm2\%$ for Oe
and $27\pm2\%$ for Be) to be double that of the LMC ($5\pm1\%$ for Oe
and $16\pm2\%$ for Be), when including the ``photometric'' Oe and Be
stars. This is the first time the frequency of Oe stars is determined
beyond the Galaxy and at subsolar metallicities. We also find Be stars
and Be/X-ray binaries to occur at higher luminosity, sgB[e] stars to be
on average less luminous than their counterparts in the LMC (at [3.6]),
and WR stars to have smaller excess, all due to the different evolution
at the lower metallicity of the SMC.

The infrared colors of the RSGs in the SMC are found in most cases to be
consistent with little dust, with only the most luminous sources showing
excess emission presumably from circumstellar dust. This is in contrast
to the generally dusty RSGs in the LMC, and agrees with the expectation
that the dust content in metal-poor RSGs is lower. We find that the
mass-loss rates in SMC RSGs correlate positively with luminosity, as in
the LMC. Finally, we confirm the astrophysical peculiarity of the
composite A \& F type spectra discovered by \citet{Evans04}, the sgB[e]
nature of 2dFS1804, and find the F0 supergiant 2dFS3528 to be an LBV
candidate. This paper thus demonstrates the wealth of information
contained in the SAGE-SMC survey, and enables studies of the infrared
properties of massive stars as a function of metallicity, in combination
with the SAGE survey of the LMC. Studies of particular classes of
massive stars and whole massive star populations in nearby galaxies, as
outlined in Paper I, are obvious directions for follow-up.

\acknowledgments{AZB acknowledges support from the Riccardo Giacconi
  Fellowship award of the Space Telescope Science Institute and from the
  European Commission Framework Program Seven under a Marie Curie
  International Reintegration Grant. The {\it Spitzer} SAGE-SMC project
  was supported by NASA grant NAG5-12595. This work is based [in part]
  on archival data obtained with the {\it Spitzer} Space Telescope,
  which is operated by the Jet Propulsion Laboratory, California
  Institute of Technology under a contract with NASA. Support for this
  work was provided by an award issued by JPL/Caltech. This publication
  makes use of data products from the 2MASS, which is a joint project of
  the University of Massachusetts and the Infrared Processing and
  Analysis Center/California Institute of Technology, funded by the
  National Aeronautics and Space Administration and the National Science
  Foundation.}

{\it Facility:} \facility{Spitzer (IRAC, MIPS)}

\clearpage

\begin{deluxetable}{lclcl}
\tabletypesize{\footnotesize}
\tablewidth{0pc}
\tablecaption{\sc Catalog of Spectral Types for 5324 SMC Massive Stars}
\tablehead{\colhead{Star} & \colhead{RA (J2000)} &
\colhead{Dec (J2000)} & \colhead{Reference$^{b}$} & \colhead{Classification}  \\
\colhead{Name$^{a}$} & \colhead{(deg)} &
\colhead{(deg)} & \colhead{} & \colhead{\& Comments}}
\startdata
2dFS0001  &  6.944625  &  $-$73.27911  &        E04  &                 O9.5:  \\
2dFS0002  &  7.036375  &  $-$73.49611  &        E04  &                 B0-5V  \\
2dFS0003  &  7.089875  &  $-$73.17531  &        E04  &                 B0-5V  \\
2dFS0004  &  7.102458  &  $-$73.88650  &        E04  &               B0-5III  \\
2dFS0005  &  7.137542  &  $-$73.30739  &        E04  &                 B0-5V  \\
2dFS0006  &  7.243000  &  $-$73.46847  &        E04  &                  A2II  \\
2dFS0007  &  7.246125  &  $-$73.28139  &        E04  &                B1-3II  \\
2dFS0008  &  7.322250  &  $-$73.14731  &        E04  &                B0-5IV  \\
\enddata                         
\label{tab:catalog}
\tablenotetext{a}{Star designations: \citet[2dFS;][]{Evans04},
  \citet[AzV;][]{Azzopardi75, Azzopardi79, Azzopardi82},
  \citet[R;][]{Feast60}, \citet[Sk;][]{Sanduleak68,Sanduleak69},
  \citet[MPG;][]{Massey89}.}

\tablenotetext{b}{Reference key: (A09) \citet{Antoniou09}, (B95)
  \citet{Barba95}, (B01) \citet{Bartzakos01}, (F03) \citet{Foellmi03},
  (E04) \citet{Evans04}, (E06) \citet{Evans06}, (E10) \citet{Hunter10},
  (GC87) \citet{Garmany87}, (H83) \citet{Humphreys83}, (H03)
  \citet{Harries03}, (H05) \citet{Hilditch05}, (L97) \citet{Lennon97},
  (L06) \citet{Levesque06}, (L07) \citet{Levesque07}, (MPG89)
  \citet{Massey89}, (M95) \citet{Massey95}, (M00) \citet{Massey00},
  (M02) \citet{Massey02}, (MO03) \citet{Massey03b}, (M03)
  \citet{Massey03c}, (M05) \citet{Massey05}, (M06) \citet{Mennickent06},
  (M07) \citet{Martayan07a,Martayan07b}, (M09) \citet{Massey09a}, (P08)
  \citet{Prieto08c}, (R05) \citet{Raguzova05}, (SB97)
  \citet{SmithNeubig97}, (W07) \citet{Wisniewski07}, (Z06)
  \citet{Zickgraf06}.}

\tablecomments{Table~\ref{tab:catalog} is available in its entirety in the
electronic version of the Journal. A portion is shown here for guidance
regarding its form and content.}
\end{deluxetable} 

\begin{deluxetable}{lr}
\tabletypesize{\footnotesize}
\tablewidth{0pc}
\tablecaption{\sc Statistics for the 3654 Matched Stars}
\tablehead{\colhead{Catalogs Matched} & \colhead{\# Stars} }

\startdata
IRACC	                    &   40  \\
IRACC+IRSF	            &  214 \\
IRACC+IRSF+OGLE	            &  352 \\
IRACC+MCPS	            &    3 \\
IRACC+MCPS+IRSF	            &   57  \\
IRACC+MCPS+IRSF+OGLE	    & 2733 \\
IRACC+MCPS+OGLE	            &   54 \\
IRACC+MCPS+MIPS24           &    3 \\
IRACC+MIPS24                &    6 \\
IRACC+MIPS24+IRSF	    &   35 \\
IRACC+MIPS24+IRSF+OGLE	    &   10 \\
IRACC+MIPS24+MCPS+IRSF	    &   80 \\
IRACC+MIPS24+MCPS+IRSF+OGLE &   32 \\
IRACC+OGLE	            &   35 \\
\enddata                         
\label{tab:matchtype}
\end{deluxetable} 

\begin{deluxetable}{cllllllllll}
\tabletypesize{\footnotesize}
\tablewidth{0pc}
\tablecaption{\sc 0.3-24 $\mu${\rm m} Photometry of 3654 Massive Stars in the SMC}
\tablehead{\colhead{Star Name$^a$} & \colhead{IRAC Designation} &
\colhead{RA(J2000)} & \colhead{Dec(J2000)} &
\colhead{$U$} & \colhead{$\sigma_U$} &
\colhead{$B$} & \colhead{$\sigma_B$} &\colhead{...} &
\colhead{Ref.$^b$} & \colhead{Classification \& Comments}}
\startdata
2dFS0005 & J002832.82$-$731827.0 & 7.137542 & $-$73.30739 & 16.33  & 0.042 & 17.063 & 0.023  & ... &E04 & B0$-$5V \\ 
2dFS0006 & J002858.15$-$732807.0 & 7.243    & $-$73.46848 & 16.648 & 0.048 & 16.605 & 0.056  & ... &E04 & A2II  \\ 
2dFS0010 & J002923.69$-$734424.6 & 7.349375 & $-$73.73997 & 16.706 & 0.041 & 17.318 & 0.033  & ... &E04 & B0$-$5V \\ 
2dFS0016 & J003033.08$-$735709.9 & 7.6385   & $-$73.95272 & 15.977 & 0.052 & 16.051 & 0.03   & ... &E04 & A0II  \\ 
2dFS0019 & J003049.36$-$740000.3 & 7.706292 & $-$73.99995 & 16.006 & 0.05  & 16     & 0.041  & ... &E04 & A0II  \\ 
2dFS0024 & J003107.03$-$732709.9 & 7.780042 & $-$73.45267 & 16.611 & 0.114 & 17.119 & 0.034  & ... &E04 & B3IV  \\ 
2dFS0026 & J003111.52$-$735936.1 & 7.798458 & $-$73.99333 & 15.744 & 0.049 & 16.292 & 0.033  & ... &E04 & B1$-$3V \\ 
\enddata                         
\label{tab:phot}
\tablenotetext{a}{Star designations: \citet[2dFS;][]{Evans04},
  \citet[AzV;][]{Azzopardi75, Azzopardi79, Azzopardi82},
  \citet[R;][]{Feast60}, \citet[Sk;][]{Sanduleak68,Sanduleak69},
  \citet[MPG;][]{Massey89}.}

\tablenotetext{b}{Reference key: (B95) \citet{Barba95}, (B01)
  \citet{Bartzakos01}, (E04) \citet{Evans04}, (E06) \citet{Evans06},
  (E10) \citet{Hunter10}, (F03) \citet{Foellmi03}, (GC87)
  \citet{Garmany87}, (H83) \citet{Humphreys83}, (H03) \citet{Harries03},
  (H05) \citet{Hilditch05}, (L97) \citet{Lennon97}, (L06)
  \citet{Levesque06}, (L07) \citet{Levesque07}, (MPG89)
  \citet{Massey89}, (M95) \citet{Massey95}, (M02) \citet{Massey02},
  (MO03) \citet{Massey03b}, (M06) \citet{Mennickent06}, (M07)
  \citet{Martayan07a,Martayan07b}, (M09) \citet{Massey09a}, (P08)
  \citet{Prieto08c}, (R05) \citet{Raguzova05}, (SB97)
  \citet{SmithNeubig97}, (W07) \citet{Wisniewski07}, (Z06)
  \citet{Zickgraf06}.}

\tablecomments{Table~\ref{tab:phot} is available in its entirety in the
electronic version of the Journal. A portion is shown here for guidance
regarding its form and content.}
\end{deluxetable} 

\begin{deluxetable}{lcccc}
\tabletypesize{\footnotesize}
\tablewidth{0pc}
\tablecaption{\sc Filter \& Detection Characteristics}
\tablehead{\colhead{Filter}  & \colhead{$\lambda_{\rm eff}$}  &
\colhead{Zero mag} & \colhead{Resolution} & \colhead{\# stars}\\
\colhead{}  & \colhead{($\mu$m)}  &
\colhead{flux (Jy)} & \colhead{($\arcsec$)} & \colhead{detected}}
\startdata
$U$  &        0.36  &    1790 & 1.5/2.6 &  2921   \\
$B$ & 	      0.44  &    4063 & 1.5/2.6 &  2962   \\
$V$ & 	      0.555  &    3636 & 1.5/2.6 &  2962   \\
$I$ & 	      0.79  &    2416 & 1.5/2.6 &  2844   \\
$V_{OGLE}$ &  0.555  &    3636 & 1.2 &  3152    \\
$I_{OGLE}$ &  0.79  &    2416 & 1.2 &   3166  \\
$J$ & 	      1.235   &   1594  & 2.5 &  3039   \\
$H$ & 	      1.662   &   1024  & 2.5 &  3007   \\
$K_s$ &       2.159   &   666.7  & 2.5 &  2789   \\
$J_{IRSF}$ &  1.235   &   1594  & 1.3 &  3449  \\
$H_{IRSF}$ &  1.662   &   1024  & 1.3 &  3413   \\
$Ks_{IRSF}$ & 2.159   &   666.7  & 1.3 &  3345   \\
$[3.6]$ &     3.55   &   280.9  & 1.7 &   3509  \\
$[4.5]$ &     4.493   &   179.7  & 1.7 &   3111  \\
$[5.8]$ &     5.731   &   115.0  & 1.9 &  1038  \\
$[8.0]$ &     7.872   &   64.13  & 2 &   509  \\
$[24]$ &      23.68   &   7.14  & 6 & 166   \\
\enddata                         
\label{tab:filters}

\end{deluxetable} 

\begin{deluxetable}{cll}
\tabletypesize{\footnotesize} \tablewidth{0pc} \tablecaption{Oe \sc and
  \rm Be \sc Star Fractions} \tablehead{\colhead{Fraction}
  &\colhead{LMC} & \colhead{SMC}} \startdata Oe/O & $1\pm0.6\%$ &
$2\pm1\%$ \\ (Oe+``phot Oe'')/O & $5\pm1\%$ & $10\pm2\%$ \\ Be/early-B &
$4\pm1\%$ & $19\pm1\%$ \\ (Be+``phot Be'')/early-B & $16\pm2\%$ &
$27\pm2\%$ \\ \enddata
\label{tab:oebefractions}

\end{deluxetable}


\clearpage
\begin{thebibliography}{100}
\expandafter\ifx\csname natexlab\endcsname\relax\def\natexlab#1{#1}\fi

\bibitem[{{Antoniou} {et~al.}(2009){Antoniou}, {Hatzidimitriou}, {Zezas}, \&
  {Reig}}]{Antoniou09}
{Antoniou}, V., {Hatzidimitriou}, D., {Zezas}, A., \& {Reig}, P. 2009, \apj,
  707, 1080

\bibitem[{{Arp}(1959)}]{Arp59}
{Arp}, B.~H. 1959, \aj, 64, 254

\bibitem[{{Azzopardi} \& {Vigneau}(1979)}]{Azzopardi79}
{Azzopardi}, M. \& {Vigneau}, J. 1979, \aaps, 35, 353

\bibitem[{{Azzopardi} \& {Vigneau}(1982)}]{Azzopardi82}
---. 1982, \aaps, 50, 291

\bibitem[{{Azzopardi} {et~al.}(1975){Azzopardi}, {Vigneau}, \&
  {Macquet}}]{Azzopardi75}
{Azzopardi}, M., {Vigneau}, J., \& {Macquet}, M. 1975, \aaps, 22, 285

\bibitem[{{Barb{\'a}} {et~al.}(1995){Barb{\'a}}, {Niemela}, {Baume},
  {et~al.}}]{Barba95}
{Barb{\'a}}, R.~H., {Niemela}, V.~S., {Baume}, G., {et~al.} 1995, \apjl, 446,
  L23

\bibitem[{{Bartzakos} {et~al.}(2001){Bartzakos}, {Moffat}, \&
  {Niemela}}]{Bartzakos01}
{Bartzakos}, P., {Moffat}, A.~F.~J., \& {Niemela}, V.~S. 2001, \mnras, 324, 18

\bibitem[{{Bolatto} {et~al.}(2007){Bolatto}, {Simon}, {Stanimirovi{\'c}},
  {et~al.}}]{Bolatto07}
{Bolatto}, A.~D., {Simon}, J.~D., {Stanimirovi{\'c}}, S., {et~al.} 2007, \apj,
  655, 212

\bibitem[{{Bonanos} {et~al.}(2009){Bonanos}, {Massa}, {Sewilo},
  {et~al.}}]{Bonanos09a}
{Bonanos}, A.~Z., {Massa}, D.~L., {Sewilo}, M., {et~al.} 2009, \aj, 138, 1003

\bibitem[{{de Jager} {et~al.}(1988){de Jager}, {Nieuwenhuijzen}, \& {van der
  Hucht}}]{deJager88}
{de Jager}, C., {Nieuwenhuijzen}, H., \& {van der Hucht}, K.~A. 1988, \aaps,
  72, 259

\bibitem[{{Dougherty} {et~al.}(1994){Dougherty}, {Waters}, {Burki},
  {et~al.}}]{Dougherty94}
{Dougherty}, S.~M., {Waters}, L.~B.~F.~M., {Burki}, G., {et~al.} 1994, \aap,
  290, 609

\bibitem[{{Dray}(2006)}]{Dray06}
{Dray}, L.~M. 2006, \mnras, 370, 2079

\bibitem[{{Ekstr{\"o}m} {et~al.}(2008){Ekstr{\"o}m}, {Meynet}, {Maeder}, \&
  {Barblan}}]{Ekstrom08}
{Ekstr{\"o}m}, S., {Meynet}, G., {Maeder}, A., \& {Barblan}, F. 2008, \aap,
  478, 467

\bibitem[{{Evans} \& {Howarth}(2008)}]{Evans08}
{Evans}, C.~J. \& {Howarth}, I.~D. 2008, \mnras, 386, 826

\bibitem[{{Evans} {et~al.}(2004{\natexlab{a}}){Evans}, {Howarth}, {Irwin},
  {et~al.}}]{Evans04}
{Evans}, C.~J., {Howarth}, I.~D., {Irwin}, M.~J., {et~al.} 2004{\natexlab{a}},
  \mnras, 353, 601

\bibitem[{{Evans} {et~al.}(2006){Evans}, {Lennon}, {Smartt},
  {et~al.}}]{Evans06}
{Evans}, C.~J., {Lennon}, D.~J., {Smartt}, S.~J., {et~al.} 2006, \aap, 456, 623

\bibitem[{{Evans} {et~al.}(2004{\natexlab{b}}){Evans}, {Lennon}, {Trundle},
  {et~al.}}]{Evans04b}
{Evans}, C.~J., {Lennon}, D.~J., {Trundle}, C., {et~al.} 2004{\natexlab{b}},
  \apj, 607, 451

\bibitem[{{Feast} {et~al.}(1960){Feast}, {Thackeray}, \& {Wesselink}}]{Feast60}
{Feast}, M.~W., {Thackeray}, A.~D., \& {Wesselink}, A.~J. 1960, \mnras, 121,
  337

\bibitem[{{Foellmi} {et~al.}(2008){Foellmi}, {Koenigsberger}, {Georgiev},
  {et~al.}}]{Foellmi08}
{Foellmi}, C., {Koenigsberger}, G., {Georgiev}, L., {et~al.} 2008, Revista
  Mexicana de Astronom{\'i}a y Astrof{\'i}sica, 44, 3

\bibitem[{{Foellmi} {et~al.}(2003){Foellmi}, {Moffat}, \&
  {Guerrero}}]{Foellmi03}
{Foellmi}, C., {Moffat}, A.~F.~J., \& {Guerrero}, M.~A. 2003, \mnras, 338, 360

\bibitem[{{Garmany} {et~al.}(1987){Garmany}, {Conti}, \& {Massey}}]{Garmany87}
{Garmany}, C.~D., {Conti}, P.~S., \& {Massey}, P. 1987, \aj, 93, 1070

\bibitem[{{Garmany} \& {Humphreys}(1985)}]{Garmany85}
{Garmany}, C.~D. \& {Humphreys}, R.~M. 1985, \aj, 90, 2009

\bibitem[{{Gordon} {et~al.}(2010){Gordon}, {Meixner}, {Blum},
  {et~al.}}]{Gordon10}
{Gordon}, K.~D., {Meixner}, M., {Blum}, R., {et~al.} 2010, AJ, in preparation

\bibitem[{{Grebel} {et~al.}(1992){Grebel}, {Richtler}, \& {de Boer}}]{Grebel92}
{Grebel}, E.~K., {Richtler}, T., \& {de Boer}, K.~S. 1992, \aap, 254, L5

\bibitem[{{Groenewegen}(2006)}]{Groenewegen06}
{Groenewegen}, M.~A.~T. 2006, \aap, 448, 181

\bibitem[{{Groenewegen} \& {Blommaert}(1998)}]{Groenewegen98}
{Groenewegen}, M.~A.~T. \& {Blommaert}, J.~A.~D.~L. 1998, \aap, 332, 25

\bibitem[{{Groenewegen} {et~al.}(2009){Groenewegen}, {Sloan}, {Soszy{\'n}ski},
  {et~al.}}]{Groenewegen09}
{Groenewegen}, M.~A.~T., {Sloan}, G.~C., {Soszy{\'n}ski}, I., {et~al.} 2009,
  \aap, 506, 1277

\bibitem[{{Habing} {et~al.}(1994){Habing}, {Tignon}, \& {Tielens}}]{Habing94}
{Habing}, H.~J., {Tignon}, J., \& {Tielens}, A.~G.~G.~M. 1994, \aap, 286, 523

\bibitem[{{Harries} {et~al.}(2003){Harries}, {Hilditch}, \&
  {Howarth}}]{Harries03}
{Harries}, T.~J., {Hilditch}, R.~W., \& {Howarth}, I.~D. 2003, \mnras, 339, 157

\bibitem[{{Hilditch} {et~al.}(2005){Hilditch}, {Howarth}, \&
  {Harries}}]{Hilditch05}
{Hilditch}, R.~W., {Howarth}, I.~D., \& {Harries}, T.~J. 2005, \mnras, 357, 304

\bibitem[{{Humphreys}(1983)}]{Humphreys83}
{Humphreys}, R.~M. 1983, \apj, 265, 176

\bibitem[{{Hunter} {et~al.}(2007){Hunter}, {Dufton}, {Smartt},
  {et~al.}}]{Hunter07}
{Hunter}, I., {Dufton}, P.~L., {Smartt}, S.~J., {et~al.} 2007, \aap, 466, 277

\bibitem[{{Hunter} {et~al.}(in preparation){Hunter}, {Evans},
  {et~al.}}]{Hunter10}
{Hunter}, I., {Evans}, C., {et~al.} in preparation

\bibitem[{{Ita} {et~al.}(2010{\natexlab{a}}){Ita}, {Matsuura}, {Ishihara},
  {et~al.}}]{Ita10b}
{Ita}, Y., {Matsuura}, M., {Ishihara}, D., {et~al.} 2010{\natexlab{a}},
A\&A, 514, 2

\bibitem[{{Ita} {et~al.}(2010{\natexlab{b}}){Ita}, {Onaka}, {Tanabe},
  {et~al.}}]{Ita10a}
{Ita}, Y., {Onaka}, T., {Tanabe}, T., {et~al.} 2010{\natexlab{b}}, PASJ,
62, 273

\bibitem[{{Kato} {et~al.}(2007){Kato}, {Nagashima}, {Nagayama},
  {et~al.}}]{Kato07}
{Kato}, D., {Nagashima}, C., {Nagayama}, T., {et~al.} 2007, \pasj, 59, 615

\bibitem[{{Kippenhahn} \& {Weigert}(1990)}]{Kippenhahn90}
{Kippenhahn}, R. \& {Weigert}, A. 1990, {Stellar Structure and Evolution}
  (Springer-Verlag Berlin Heidelberg New York.)

\bibitem[{{Lanz} \& {Hubeny}(2003)}]{Lanz03}
{Lanz}, T. \& {Hubeny}, I. 2003, \apjs, 146, 417

\bibitem[{{Lanz} \& {Hubeny}(2007)}]{Lanz07}
---. 2007, \apjs, 169, 83

\bibitem[{{Larsen} {et~al.}(2000){Larsen}, {Clausen}, \& {Storm}}]{Larsen00}
{Larsen}, S.~S., {Clausen}, J.~V., \& {Storm}, J. 2000, \aap, 364, 455

\bibitem[{{Leitherer} {et~al.}(1992){Leitherer}, {Robert}, \&
  {Drissen}}]{Leitherer92}
{Leitherer}, C., {Robert}, C., \& {Drissen}, L. 1992, \apj, 401, 596

\bibitem[{{Lennon}(1997)}]{Lennon97}
{Lennon}, D.~J. 1997, \aap, 317, 871

\bibitem[{{Lennon} {et~al.}(1993){Lennon}, {Mazzali}, {Pasian},
  {et~al.}}]{Lennon93}
{Lennon}, D.~J., {Mazzali}, P.~A., {Pasian}, F., {et~al.} 1993, Space Science
  Reviews, 66, 169

\bibitem[{{Levesque} {et~al.}(2006){Levesque}, {Massey}, {Olsen},
  {et~al.}}]{Levesque06}
{Levesque}, E.~M., {Massey}, P., {Olsen}, K.~A.~G., {et~al.} 2006, \apj, 645,
  1102

\bibitem[{{Levesque} {et~al.}(2007){Levesque}, {Massey}, {Olsen},
  {et~al.}}]{Levesque07}
---. 2007, \apj, 667, 202

\bibitem[{{Liu} {et~al.}(2005){Liu}, {van Paradijs}, \& {van den
  Heuvel}}]{Liu05}
{Liu}, Q.~Z., {van Paradijs}, J., \& {van den Heuvel}, E.~P.~J. 2005, \aap,
  442, 1135

\bibitem[{{Maeder} {et~al.}(1999){Maeder}, {Grebel}, \&
  {Mermilliod}}]{Maeder99}
{Maeder}, A., {Grebel}, E.~K., \& {Mermilliod}, J. 1999, \aap, 346, 459

\bibitem[{{Marshall} {et~al.}(2004){Marshall}, {van Loon}, {Matsuura}, {Wood},
  {Zijlstra}, \& {Whitelock}}]{Marshall04}
{Marshall}, J.~R., {van Loon}, J.~T., {Matsuura}, M., {Wood}, P.~R.,
  {Zijlstra}, A.~A., \& {Whitelock}, P.~A. 2004, \mnras, 355, 1348

\bibitem[{{Martayan} {et~al.}(2010){Martayan}, {Baade}, \&
  {Fabregat}}]{Martayan10}
{Martayan}, C., {Baade}, D., \& {Fabregat}, J. 2010, \aap, 509, A11

\bibitem[{{Martayan} {et~al.}(2007{\natexlab{a}}){Martayan}, {Floquet},
  {Hubert}, {et~al.}}]{Martayan07a}
{Martayan}, C., {Floquet}, M., {Hubert}, A.~M., {et~al.} 2007{\natexlab{a}},
  \aap, 472, 577

\bibitem[{{Martayan} {et~al.}(2007{\natexlab{b}}){Martayan}, {Fr{\'e}mat},
  {Hubert}, {et~al.}}]{Martayan07b}
{Martayan}, C., {Fr{\'e}mat}, Y., {Hubert}, A., {et~al.} 2007{\natexlab{b}},
  \aap, 462, 683

\bibitem[{{Massey}(2002)}]{Massey02}
{Massey}, P. 2002, \apjs, 141, 81

\bibitem[{{Massey}(2009)}]{Massey09}
---. 2009, arXiv:0903.0155

\bibitem[{{Massey} {et~al.}(1995){Massey}, {Lang}, {Degioia-Eastwood},
  {et~al.}}]{Massey95}
{Massey}, P., {Lang}, C.~C., {Degioia-Eastwood}, K., {et~al.} 1995, \apj, 438,
  188

\bibitem[{{Massey} \& {Olsen}(2003)}]{Massey03b}
{Massey}, P. \& {Olsen}, K.~A.~G. 2003, \aj, 126, 2867

\bibitem[{{Massey} {et~al.}(2003){Massey}, {Olsen}, \& {Parker}}]{Massey03c}
{Massey}, P., {Olsen}, K.~A.~G., \& {Parker}, J.~W. 2003, \pasp, 115, 1265

\bibitem[{{Massey} {et~al.}(1989){Massey}, {Parker}, \& {Garmany}}]{Massey89}
{Massey}, P., {Parker}, J.~W., \& {Garmany}, C.~D. 1989, \aj, 98, 1305

\bibitem[{{Massey} {et~al.}(2005){Massey}, {Puls}, {Pauldrach},
  {et~al.}}]{Massey05}
{Massey}, P., {Puls}, J., {Pauldrach}, A.~W.~A., {et~al.} 2005, \apj, 627, 477

\bibitem[{{Massey} {et~al.}(2000){Massey}, {Waterhouse}, \&
  {DeGioia-Eastwood}}]{Massey00}
{Massey}, P., {Waterhouse}, E., \& {DeGioia-Eastwood}, K. 2000, \aj, 119, 2214

\bibitem[{{Massey} {et~al.}(2009){Massey}, {Zangari}, {Morrell},
  {et~al.}}]{Massey09a}
{Massey}, P., {Zangari}, A.~M., {Morrell}, N.~I., {et~al.} 2009, \apj, 692, 618

\bibitem[{{McDonald} {et~al.}(2009){McDonald}, {van Loon}, {Decin},
  {et~al.}}]{McDonald09}
{McDonald}, I., {van Loon}, J.~T., {Decin}, L., {et~al.} 2009, \mnras, 394, 831

\bibitem[{{Meixner} {et~al.}(2006){Meixner}, {Gordon}, {Indebetouw},
  {et~al.}}]{Meixner06}
{Meixner}, M., {Gordon}, K.~D., {Indebetouw}, R., {et~al.} 2006, \aj, 132, 2268

\bibitem[{{Mennickent} {et~al.}(2006){Mennickent}, {Cidale}, {Pietrzy{\'n}ski},
  {et~al.}}]{Mennickent06}
{Mennickent}, R.~E., {Cidale}, L., {Pietrzy{\'n}ski}, G., {et~al.} 2006, \aap,
  457, 949

\bibitem[{{Mokiem} {et~al.}(2007){Mokiem}, {de Koter}, {Vink},
  {et~al.}}]{Mokiem07a}
{Mokiem}, M.~R., {de Koter}, A., {Vink}, J.~S., {et~al.} 2007, \aap, 473, 603

\bibitem[{{Negueruela} {et~al.}(2004){Negueruela}, {Steele}, \&
  {Bernabeu}}]{Negueruela04}
{Negueruela}, I., {Steele}, I.~A., \& {Bernabeu}, G. 2004, Astronomische
  Nachrichten, 325, 749

\bibitem[{{Oestreicher} {et~al.}(1997){Oestreicher}, {Schmidt-Kaler}, \&
  {Wargau}}]{Oestreicher97}
{Oestreicher}, M.~O., {Schmidt-Kaler}, T., \& {Wargau}, W. 1997, \mnras, 289,
  729

\bibitem[{{Osmer}(1973)}]{Osmer73}
{Osmer}, P.~S. 1973, \apj, 181, 327

\bibitem[{{Panagia} \& {Felli}(1975)}]{Panagia75}
{Panagia}, N. \& {Felli}, M. 1975, \aap, 39, 1

\bibitem[{{Panagia} \& {Thum}(1981)}]{Panagia81}
{Panagia}, N. \& {Thum}, C. 1981, \aap, 98, 295

\bibitem[{{Pojma\'nski}(2002)}]{Pojmanski02}
{Pojma\'nski}, G. 2002, Acta Astronomica, 52, 397

\bibitem[{{Porter} \& {Rivinius}(2003)}]{Porter03a}
{Porter}, J.~M. \& {Rivinius}, T. 2003, \pasp, 115, 1153

\bibitem[{{Prieto} {et~al.}(2008){Prieto}, {Stanek}, {Kochanek},
  {et~al.}}]{Prieto08c}
{Prieto}, J.~L., {Stanek}, K.~Z., {Kochanek}, C.~S., {et~al.} 2008, \apjl, 673,
  L59

\bibitem[{{Raguzova} \& {Popov}(2005)}]{Raguzova05}
{Raguzova}, N.~V. \& {Popov}, S.~B. 2005, Astronomical and Astrophysical
  Transactions, 24, 151

\bibitem[{{Reach} {et~al.}(2005){Reach}, {Megeath}, {Cohen},
  {et~al.}}]{Reach05}
{Reach}, W.~T., {Megeath}, S.~T., {Cohen}, M., {et~al.} 2005, \pasp, 117, 978

\bibitem[{{Sanduleak}(1968)}]{Sanduleak68}
{Sanduleak}, N. 1968, \aj, 73, 246

\bibitem[{{Sanduleak}(1969)}]{Sanduleak69}
---. 1969, \aj, 74, 877

\bibitem[{{Skrutskie} {et~al.}(2006){Skrutskie}, {Cutri}, {Stiening},
  {et~al.}}]{Skrutskie06}
{Skrutskie}, M.~F., {Cutri}, R.~M., {Stiening}, R., {et~al.} 2006, \aj, 131,
  1163

\bibitem[{{Sloan} {et~al.}(2008){Sloan}, {Kraemer}, {Wood}, {et~al.}}]{Sloan08}
{Sloan}, G.~C., {Kraemer}, K.~E., {Wood}, P.~R., {et~al.} 2008, \apj, 686, 1056

\bibitem[{{Smith Neubig} \& {Bruhweiler}(1997)}]{SmithNeubig97}
{Smith Neubig}, M.~M. \& {Bruhweiler}, F.~C. 1997, \aj, 114, 1951

\bibitem[{{Srinivasan} {et~al.}(2009){Srinivasan}, {Meixner}, {Leitherer},
  {et~al.}}]{Srinivasan09}
{Srinivasan}, S., {Meixner}, M., {Leitherer}, C., {et~al.} 2009, \aj, 137, 4810

\bibitem[{{Szczygiel} {et~al.}(2010){Szczygiel}, {Stanek}, {Bonanos},
  {et~al.}}]{Szczygiel10}
{Szczygiel}, D.~M., {Stanek}, K.~Z., {Bonanos}, A.~Z., {et~al.} 2010,
AJ, 140, 14

\bibitem[{{Szyma{\'n}ski}(2005)}]{Szymanski05}
{Szyma{\'n}ski}, M.~K. 2005, Acta Astronomica, 55, 43

\bibitem[{{Udalski} {et~al.}(1997){Udalski}, {Kubiak}, \&
  {Szyma{\'n}ski}}]{Udalski97}
{Udalski}, A., {Kubiak}, M., \& {Szyma{\'n}ski}, M. 1997, Acta Astronomica, 47,
  319

\bibitem[{{Udalski} {et~al.}(2008){Udalski}, {Soszy{\'n}ski}, {Szyma{\'n}ski},
  {et~al.}}]{Udalski08b}
{Udalski}, A., {Soszy{\'n}ski}, I., {Szyma{\'n}ski}, M.~K., {et~al.} 2008, Acta
  Astronomica, 58, 329

\bibitem[{{van Loon}(2000)}]{vanLoon00}
{van Loon}, J.~T. 2000, \aap, 354, 125

\bibitem[{{van Loon}(2006)}]{vanLoon06}
{van Loon}, J.~T. 2006, in Astronomical Society of the Pacific Conference
  Series, Vol. 353, Stellar Evolution at Low Metallicity: Mass Loss,
  Explosions, Cosmology, ed. {H.~J.~G.~L.~M.~Lamers, N.~Langer, T.~Nugis, \&
  K.~Annuk}, 211

\bibitem[{{van Loon}(2007)}]{vanLoon07}
{van Loon}, J.~T. 2007, in Astronomical Society of the Pacific Conference
  Series, Vol. 378, Why Galaxies Care About AGB Stars: Their Importance as
  Actors and Probes, ed. {F.~Kerschbaum, C.~Charbonnel, \& R.~F.~Wing}, 227

\bibitem[{{van Loon} {et~al.}(2005){van Loon}, {Cioni}, {Zijlstra},
  {et~al.}}]{vanLoon05}
{van Loon}, J.~T., {Cioni}, M., {Zijlstra}, A.~A., {et~al.} 2005, \aap, 438,
  273

\bibitem[{{van Loon} {et~al.}(2008){van Loon}, {Cohen}, {Oliveira},
  {et~al.}}]{vanLoon08}
{van Loon}, J.~T., {Cohen}, M., {Oliveira}, J.~M., {et~al.} 2008, \aap, 487,
  1055

\bibitem[{{van Loon} {et~al.}(1999){van Loon}, {Groenewegen}, {de Koter},
  {et~al.}}]{vanLoon99}
{van Loon}, J.~T., {Groenewegen}, M.~A.~T., {de Koter}, A., {et~al.} 1999,
  \aap, 351, 559

\bibitem[{{van Loon} {et~al.}(2010{\natexlab{a}}){van Loon}, {Oliveira},
  {Gordon}, {et~al.}}]{vanLoon10a}
{van Loon}, J.~T., {Oliveira}, J.~M., {Gordon}, K.~D., {et~al.}
  2010{\natexlab{a}}, \aj, 139, 68

\bibitem[{{van Loon} {et~al.}(2010{\natexlab{b}}){van Loon}, {Oliveira},
  {Gordon}, {et~al.}}]{vanLoon10b}
---. 2010{\natexlab{b}}, \aj, 139, 1553

\bibitem[{{Verhoelst} {et~al.}(2009){Verhoelst}, {van der Zypen}, {Hony},
  {et~al.}}]{Verhoelst09}
{Verhoelst}, T., {van der Zypen}, N., {Hony}, S., {et~al.} 2009, \aap, 498, 127

\bibitem[{{Vink} {et~al.}(2001){Vink}, {de Koter}, \& {Lamers}}]{Vink01}
{Vink}, J.~S., {de Koter}, A., \& {Lamers}, H.~J.~G.~L.~M. 2001, \aap, 369, 574

\bibitem[{{Wisniewski} \& {Bjorkman}(2006)}]{Wisniewski06}
{Wisniewski}, J.~P. \& {Bjorkman}, K.~S. 2006, \apj, 652, 458

\bibitem[{{Wisniewski} {et~al.}(2007){Wisniewski}, {Bjorkman}, {Bjorkman},
  {et~al.}}]{Wisniewski07}
{Wisniewski}, J.~P., {Bjorkman}, K.~S., {Bjorkman}, J.~E., {et~al.} 2007, \apj,
  670, 1331

\bibitem[{{Wood} {et~al.}(1992){Wood}, {Whiteoak}, {Hughes}, {et~al.}}]{Wood92}
{Wood}, P.~R., {Whiteoak}, J.~B., {Hughes}, S.~M.~G., {et~al.} 1992, \apj, 397,
  552

\bibitem[{{Wright} \& {Barlow}(1975)}]{Wright75}
{Wright}, A.~E. \& {Barlow}, M.~J. 1975, \mnras, 170, 41

\bibitem[{{Zaritsky} {et~al.}(2002){Zaritsky}, {Harris}, {Thompson},
  {et~al.}}]{Zaritsky02}
{Zaritsky}, D., {Harris}, J., {Thompson}, I.~B., {et~al.} 2002, \aj, 123, 855

\bibitem[{{Zickgraf}(2006)}]{Zickgraf06}
{Zickgraf}, F.-J. 2006, in Astronomical Society of the Pacific Conference
  Series, Vol. 355, Stars with the B[e] Phenomenon, ed. M.~{Kraus} \& A.~S.
  {Miroshnichenko}, 135

\end{thebibliography}
\end{document}